\documentclass[a4paper,11pt]{article}
\pdfoutput=1 

\usepackage{jheppub} 

\usepackage[T1]{fontenc} 

\usepackage{siunitx} 

\title{\boldmath The ChPT: top-down and bottom-up}

\author{Karol Kampf}

\affiliation{Institute of Particle and Nuclear Physics, Charles University\\V Hole\v{s}ovi\v{c}k\'{a}ch 2, 180 00 Prague 8, Czech Republic}

\emailAdd{karol.kampf@mff.cuni.cz}

\abstract{In this work, higher-derivative corrections of the non-linear sigma model of both even and odd intrinsic-parity sectors are systematically studied, focusing on ordered amplitudes of flavor scalars in massless limit. It should correspond to a theory known as chiral perturbation theory (ChPT) without external sources and with only single-trace operators. We briefly overview its formal development and apply new $S$-matrix methods to its amplitude constructions. The bottom-up analysis of the tree-level amplitudes of different orders and multiplicities focuses on the formal structure of general ChPT. Possible theoretical simplifications based on the Kleiss-Kuijf and Bern-Carrasco-Johansson relations are presented. Finally, in the same context, the comparison with the so-called $Z$-function, which is connected with string theory, is also discussed.}

\begin{document} 
\maketitle
\flushbottom

\section{Introduction}
Chiral perturbation theory (ChPT) is a low energy effective field theory of quantum chromodynamics (QCD). It describes the dynamics of the Goldstone boson particles based on the symmetry pattern of QCD: $SU(N_f)_L\times SU(N_f)_R/SU(N_F)_V$.
Ignoring the $\theta$ parameter of QCD, assuming the confinement and $N_f$ massless quarks with $N_f\geq3$ we have this spontaneous symmetry pattern as an inevitable consequence of QCD \cite{Vafa:1983tf,tHooft:1980xss}. Masses of light quarks are not zero, and this fact should be taken into account by some model assumptions and included in the chiral theory as corrections. The most natural choice is to assume that masses of Goldstone bosons are of the same order of magnitude as momenta. However, this canonical choice is not the only option, and different possibilities were also studied and confronted with experiments \cite{Knecht:1994zb,Knecht:1995tr,Descotes-Genon:2003qye,Kolesar:2008jr,Kolesar:2016jwe,Kampf:2009tk}. In this work we will completely disregard this problem as we will be working in the so-called chiral limit, i.e. we assume that $N_f$ quarks are strictly massless. It means that the Goldstone bosons are in fact genuine Goldstone bosons with zero masses.

From a top-down perspective, ChPT is represented by Lagrangian monomials whose form is dictated by the symmetry pattern of QCD.  
Without considering masses, the monomials are arranged in a series with respect to the number of derivatives. Using the Feynman rules, one can calculate any amplitude where the derivatives are translated to momenta. Studying the low energy phenomena means that orders are tight with derivatives, starting with the leading order (LO) with two derivatives (due to the Lorentz invariance) or $O(p^2)$ order in amplitudes. The non-linear sigma model (NLSM) is usually another name for this LO Lagrangian. Here we want to study the NLSM theory, including also its higher-derivative corrections. To stress this fact, we will call the theory ChPT to emphasize that we want to go beyond the leading order (i.e., beyond NLSM). This effective field theory concerns a very active field of research, spanned over several decades, studied from both the theoretical and phenomenological points of view. It started especially with Weinberg's work \cite{Weinberg:1978kz} and was considerably extended by Leutwyler and Gasser in \cite{Gasser:1984gg} where the basis for the next-to-leading order (NLO) was introduced. The number of independent monomials rises with every order, and every such term has its own low-energy constant. As only the above-mentioned symmetry pattern is used in the construction of monomials of the effective Lagrangian, the constants that in principle are dictated by the underlying QCD are taken independent and a priori unknown. Their values must be set from phenomenology \cite{Bijnens:2014lea} or by lattice simulations \cite{FlavourLatticeAveragingGroup:2019iem}.  However, one important consequence of the non-linear realization of the symmetry is non-trivially encoded in the monomials. It is the single soft limit (so-called Adler zero) of amplitudes.

We want to revisit the subject of formal expansion of ChPT with several simplifications. The first substantial reduction -- the chiral limit -- was already mentioned. Another simplification is to focus only on the Goldstone bosons and disregard any other particles (like photons, resonances, fermions, etc.). In the language of ChPT, it also means that we will not consider any external current. Last but not least, we will focus only on the single trace operators. This is a crucial simplification for us, motivated by the large $N_C$ assumptions. It enables us to perform a key technical manipulation -- stripping of the amplitude. It has two advantages -- we get rid of the group (flavor) indices, and the numbers of diagrams of amplitudes are much smaller as we have to deal only with cyclical permutations. This primitive amplitude (called ordered or stripped here) will be the main object of this work.  Apart from these simplifications, we will not disregard the chiral anomaly. It means we will include its effect in the form of the  Wess-Zumino-Witten term, which describes the odd vertices at $O(p^4)$ and include also higher orders in the odd sector.

In the first part of this work, we will merely summarize existing results in literature from the top-down perspective simplified by our assumptions. It would include mainly the list of Lagrangians at different orders. Using the standard Feynman rules, we can calculate cyclically ordered amplitudes of any multiplicity. 
The scattering amplitudes also represent the connection with our second part, where we will study them directly using the on-shell methods.

The on-shell scattering methods represent a bottom-up approach of studying the amplitudes differently, in some sense more directly, without explicit Lagrangian and Feynman rules. In recent years, this revival interest in the $S$-matrix program has brought substantial progress, especially for gauge theories and gravity. There have been many revelations in our understanding of the structure and properties of scattering amplitudes in quantum field theory.
This includes the unexpected simplicity of final results and hidden symmetries invisible in the conventional
formulations and led to many new discoveries as: generalized unitarity, recursion relations, CHY formulations, color-kinematics duality, geometrical language of amplitudes \cite{Bern:1994zx,Britto:2005fq,Cachazo:2013iea,Bern:2010ue,Bern:2019prr,Arkani-Hamed:2013jha}.
Curiously the on-shell methods are relevant also for effective field theories. In addition to NLSM or ChPT it involves recent studies within theories as DBI, Volkov-Akulov, (special) Galileons, Born-Infeld and their various combinations \cite{Kampf:2012fn,Kampf:2013vha,Cachazo:2014xea,Kampf:2014rka,Cheung:2014dqa,Cheung:2015ota,Cheung:2016drk,Cheung:2018oki,Elvang:2018dco,Bijnens:2019eze,Kampf:2019mcd,Low:2019wuv,Kampf:2020tne,Cheung:2020qxc,Low:2020ubn,Kampf:2021bet,Kampf:2021tbk}.
The on-shell scattering methods will also be employed in the second part of this work. We will use them to set the most general basis for given order with a given number of scattering particles. Such a general basis will then be used to calculate the so-called ordered amplitudes. A crucial test of Adler zero fixes the final form of the studied amplitude and consequently the final form of the basis. This basis can then be compared with existing top-down formulations.
Similar strategy was used in \cite{Dai:2020cpk, Low:2019ynd,Carrasco:2019qwr,Graf:2020yxt}. 

The Adler zero seems too little for what is possible to inherit from QCD. In fact, having established many stripped amplitudes (with different orders and different multiplicities), it is clear that they represent a vast playground for further exploration.  For example in~\cite{Chen:2013fya} it was shown that the NLSM amplitudes satisfy Kleiss-Kuijf~\cite{Kleiss:1988ne} and also Bern-Carrasco-Johansson (BCJ) relations \cite{Bern:2008qj}. These relations were originally studied for Yang-Mills theory using both the string and field theory methods. Interestingly, they are valid for the tree-level QCD amplitudes (with or without quark masses) \cite{Johansson:2015oia,delaCruz:2015dpa}.
It might be a coincidence, but it is important to study the consequences also for higher orders and define a core of ChPT satisfying KK and/or BCJ relations.
Recently, application of the BCJ up to $O(p^4)$ was done in \cite{Rodina:2021isd} and dedicated study to five points in \cite{Carrasco:2021ptp}. Both even and odd sectors up to $O(p^6)$ and up to 6pt were also studied in~\cite{CarrilloGonzalez:2019fzc}.
In the same spirit, we will also use the results of the open superstring. This theory can be generated by the double-copy of maximally supersymmetric Yang-Mills and doubly-ordered scalars called $Z$-theory \cite{Carrasco:2016ldy}. The crucial for our work is the $Z$-function, which describes the amplitudes of the scalars and curiously satisfies KK and BCJ relations. It carries the sole parameter $\alpha'$ of the superstring theory, and in the lowest limit when $\alpha'\to 0$ it coincides with the NLSM amplitudes. We will study in this work higher $\alpha'$ corrections and connection with the established ChPT+BCJ theory, i.e., a theory that fulfills both the Adler zero and BCJ constraints.

\section{Top-down: Conventional method}
In this section, we will mainly summarize the present status of the chiral Lagrangian in its two respective sectors: of the even and of the odd intrinsic parity. The construction in both sectors reflects specific symmetry properties of QCD, namely the chiral symmetry in the chiral limit (light quarks are massless), parity and charge conjugation, in combination with the general properties of quantum field theory (e.g. Hermiticity or CPT invariance). For detail discussion we refer to original papers (see below) or lectures and reviews, e.g. \cite{Pich:1998xt, Ecker:2000fe, Colangelo:2000zw, Scherer:2005ri, Kubis:2007iy, Bijnens:2006zp}.
As mentioned in the introduction we will be focusing only on the single trace operators. This enables to define the cyclically ordered stripped vertices $V_n(p_1,\ldots,p_n)$. They are connected with the standard vertices via summing over all permutations modulo cyclicity
\begin{equation}\label{eq:vertexordered}
    V_n^{a_1\ldots a_n}(p_1,\ldots,p_n) = \sum_{\sigma\in S_n/Z_n} 
    V_n (p_{\sigma(1)},\ldots, p_{\sigma(n)})\langle t^{a_{\sigma(1)}}\ldots t^{a_{\sigma(n)}} \rangle \,,
\end{equation}
where $\langle.\rangle$ represents the trace and $t^a$ are the generators of $SU(N)$ with normalization $\langle t^a t^b \rangle = \delta^{ab}$, so e.g. $t^a = \sigma^a/\sqrt2$ for $N=2$ and $t^a = \lambda^a/\sqrt2$ for $N=3$, with Pauli and Gall-Mann matrices, respectively. 

We will also briefly discuss the singlet part of the theory, i.e. for example, in the three-flavor case, the difference between the nonet and octet multiplet.

\subsection{Even intrinsic-parity sector}\label{sec:even}

Here the situation is very well explored. We will summarize and write down all relevant terms up to NNNLO as provided in the literature. Note that we are focusing only on the cyclically ordered, massless theory with the $SU(N)\times SU(N)/SU(N)$ chiral symmetry breaking pattern without external currents. The relevant terms of the Lagrangian describing dynamics of corresponding pseudoscalar Goldstone bosons start with the following leading order
\begin{equation}\label{eq:L2}
	{\cal L}_2 = \frac{F^2}{4}\langle u_\mu u^\mu \rangle\,,
\end{equation}
where we have introduced \cite{Ecker:1988te}
\begin{equation}\label{eq:def-u}
	u_\mu = i (u^\dagger\partial_\mu u - u \partial_\mu u^\dagger )\,.
\end{equation}
The most convenient exponential parametrization allows to express the $SU(N)$ multiplet of `pions' $\phi^a$ as
\begin{equation}\label{eq:udef}
	u(\phi) = \exp\left(\frac{i\Phi(\phi)}{F\sqrt2}\right),\qquad
          \Phi(\phi) = t^a\phi^a\,,\qquad a=1,\dots,N^2-1\,.
\end{equation}
For ChPT it is also common to use the so-called $LR$ basis with the building block
\begin{equation}
    U = u^2 = \exp\left(\frac{i\sqrt2\Phi(\phi)}{F}\right)\,.
\end{equation}

The NLO, or $O(p^4)$ order is given by two terms \cite{Gasser:1983yg, Gasser:1984gg}:
\begin{equation}
	{\cal L}_4 = 
		  L_0\langle u_\mu u_\nu u^\mu u^\nu \rangle 
		+ L_3\langle u_\mu u^\mu u_\nu u^\nu \rangle\,.
\end{equation}

The following higher orders are expressed conveniently using the covariant derivative and its connection
\begin{equation}
    \nabla_\mu X = \partial_\mu X + [\Gamma_\mu,X],\qquad \Gamma_\mu = \frac12(u^\dagger \partial_\mu u + u \partial_\mu u^\dagger)\,.
\end{equation}
As we consider no external currents and strict massless nature of pions, following relations hold
\begin{equation}\label{eq:relnablau}
    \nabla_\mu u_\nu - \nabla_\nu u_\mu =0,\qquad \nabla_\mu u^\mu =0,\qquad [\nabla_\mu,\nabla_\nu] u_\rho = \frac14 [[u_\mu,u_\nu],u_\rho]\,.
\end{equation}

The Lagrangian at $O(p^6)$ is represented by seven terms \cite{Bijnens:1999sh, Bijnens:2019eze} 
\begin{multline}\label{eq:L6}
    F^2 {\cal L}_6 = 
        C_1^{(6)} \langle u_\mu h^{\mu\nu}u^\rho h_{\nu\rho} \rangle
       +C_2^{(6)} \langle u_\mu h_{\nu\rho} u^\mu h^{\nu\rho} \rangle\\
       +C_3^{(6)} \langle u_\mu u^\mu u_\nu u^\nu u_\rho u^\rho \rangle
       +C_4^{(6)} \langle u_\mu u^\mu u_\nu u_\rho u^\nu u^\rho\rangle
       +C_5^{(6)} \langle u_\mu u^\mu u_\nu u_\rho u^\rho u^\nu \rangle\\
       +C_6^{(6)} \langle u_\mu u_\nu u^\mu u_\rho u^\nu u^\rho\rangle
       +C_7^{(6)} \langle u_\mu u_\nu u_\rho u^\mu u^\nu u^\rho\rangle\,.
\end{multline}
A new symbol was introduced
\begin{equation}
    h_{\mu\nu} = \nabla_\mu u_\nu + \nabla_\nu u_\mu\,.
\end{equation}
Due to the fact that $\nabla_\mu u_\nu - \nabla_\nu u_\mu$ is proportional to the external field strength and as here the external fields are zero we can freely exchange $h_{\mu\nu}$ and $\nabla_\mu u_\nu$ (modulo 2).
Note that so far we have been using original notation (constants $F$ and $L_i$), however, for $O(p^6)$ we slightly changed it to $C_i^{(6)}$. In the original paper \cite{Bijnens:1999sh} these are denoted as $K_i$. This notation will be used also for higher orders. In this logic the $O(p^4)$ constants could be relabelled as
\begin{equation}
    C_1^{(4)} = L_0\,,\qquad C_2^{(4)} = L_3\,.
\end{equation}
Note also that we want all $C_i^{(p)}$ to be dimensionless -- this is the reason for having $F^2$ in~(\ref{eq:L6}) so that only $F$ from all parameters is the dimensionfull quantity.
There is one-to-one correspondence between $C_i^{(6)}$ and $L_{6,i}$ of \cite{Bijnens:2019eze}, but only those terms with single trace are kept here:
\begin{equation}
    C_1^{(6)} = \tfrac{1}{4}L_{6,3},\quad
    C_2^{(6)} = \tfrac{1}{4}L_{6,4},\quad
    C_{3\ldots7}^{(6)} = L_{6,15\ldots 19}\,.
\end{equation}
The factor $\tfrac14$ is due to keeping $h_{\mu\nu}$ instead of $\nabla_\mu u_\nu$ in monomials.
The connection with the original $K_i$s is worked out in Appendix A of \cite{Bijnens:2019eze}.

Finally the highest available order in the literature is $O(p^8)$ completed only recently in \cite{Bijnens:2018lez}. Authors in this publication followed several guiding principles in constructing the basis, namely they removed as many terms as possible in favor of the terms with external fields and in with higher number of mesons. They also kept terms with lower number of flavor traces making the $N_C$ counting explicit. It makes the selection of the wanted monomials trivial, we have to simply read off the single-trace terms from \cite{Bijnens:2018lez}, with $u_\mu$ or $\nabla_\alpha u_\beta$ only. Schematically these are:
\begin{align}\label{eq:L8}
    F^4 {\cal L}_8 = \sum_{i=1}^3 C_i^{(8)} \langle h^4 \rangle + \sum_{i=4}^{25} C_i^{(8)} \langle h^2 u^4 \rangle + \sum_{i=26}^{42} C_i^{(8)} \langle u^8 \rangle\,,
\end{align}
for a complete form see Appendix~\ref{sec:apA}. In summary, there are three terms describing at least four pions, 22 terms starting with six pions, and finally 17 terms starting with eight pions. These numbers including also previous orders are summarized in Tab.~\ref{tab:numberofpions1}.
\begin{table}[t]
\begin{center}  
\begin{tabular}{|c|c|c|}
\hline
   &\#mesons&\#terms \\ \hline 
$p^2$ & 4 & 1  \\
\hline
$p^4$ & 4 & 2   \\
\hline
$p^6$ & 4 & 2   \\
      & 6 & 5 \\
\hline
$p^8$ & 4 & 3  \\
      & 6 & 22 \\
      & 8 & 17 \\
\hline
    \end{tabular}
   \renewcommand{\arraystretch}{1}
     
    \caption{Number of monomials that produce vertices starting at the given number of mesons.}\label{tab:numberofpions1}
  \end{center}
\end{table}
Note that no new symbol had to be introduced in (\ref{eq:L8}). Similarly as at $O(p^2)$ and $O(p^4)$ we need only $u_\mu$ at $O(p^6)$ {\it and\/} $O(p^8)$ we need only $u_\mu$ and the first covariant derivative of it ($h_{\mu\nu}$).
We can thus conjecture that at $O(p^{10})$ we will need a new symbol with two covariant derivatives
\begin{equation}\label{eq:hmnr}
    h_{\mu\nu\rho} = \nabla_\mu \nabla_\nu u_\rho + \text{sym} \,,
\end{equation}
where the symmetrization over all Lorentz indices is a consequence of~(\ref{eq:relnablau}).

\subsection{Odd intrinsic-parity sector}\label{sec:odd}

The story of the low energy limit of QCD would not be complete without studying the chiral anomaly. 
In fact, in the previous section, we have encountered an additional symmetry, namely conservation of the operation $\phi \leftrightarrow -\phi$. This is a trivial consequence of our focus only on the even number of fields. However, such a restriction is not dictated by QCD. In fact, the best-known example is the decay $\pi^0 \to \gamma\gamma$, or rather, as we are not interested in external currents, $K^+K^-\to 3 \pi$. Now, we will thus turn our attention also on the odd number of fields. The top-down solution to this problem is well known. One should start with a careful analysis of the symmetry pattern in QCD.  Using the Ward identities, i.e. the consequence of the local transformation of external fields, one has to face the axial anomaly \cite{Adler:1969er}. Solution of the inhomogeneous Ward identities thus includes apart from the general (we can refer to as chiral invariant) part also one particular solution corresponding to the anomaly. The latter is usually denoted as WZW (due to Wess, Zumino and Witten), where the part without the external fields (identified by Witten) can be written using the five-dimensional integral, i.e.~\cite{Witten:1983tw}
\begin{equation}\label{eq:L5}
    {\cal L}^A_4 = -\frac{i N_C}{48\pi^2}\int_0^1 dz \epsilon^{\mu\nu\alpha\beta}
    \langle \Sigma_z \Sigma_\mu \Sigma_\nu \Sigma_\alpha \Sigma_\beta \rangle\,,
\end{equation}
where $\Sigma_i = {\cal U}^\dagger \partial_i {\cal U}$ with ${\cal U}$ describing coset on the five dimensional bulk. The easiest parametrization is connected with the exponential one: 
\begin{equation}\label{eq:Uz}
{\cal U} = \exp \bigl( i z \sqrt{2} \Phi/F \bigr)\,.
\end{equation}
Of course, this is not the only possibility, and it would be instructive to study different parametrizations similarly to the study of LO in the even sector \cite{Kampf:2012fn, Kampf:2013vha}. 
Even though our main focus here is the amplitudes, and we are not particularly interested in `classical' calculation {\it from\/} the Lagrangian, we will perform the simplest non-trivial computation in both ways. The calculations of the seven-pion scattering from Lagrangian and using the amplitude methods were already performed in~\cite{Low:2019ynd}. We will recalculate it in another parametrization, so-called Cayley. There are two reasons. First, it is useful to perform an independent check of the calculation, as the final physical results, in this case, the striped seven-point amplitudes, have to agree in any parametrization (similarly to works of leading logs \cite{Bijnens:2012hf, Bijnens:2013yca} where several parametrizations were introduced for this purpose). The second reason is practical --  we know the vertices of~(\ref{eq:L5}) in Cayley parametrization for all multiplicities (for the proof, see Appendix~\ref{sec:cayley}).

Anomalous chiral Lagrangian at $O(p^6)$ was constructed independently some time ago in~\cite{Ebertshauser:2001nj} and~\cite{Bijnens:2001bb}. There are in total 24 monomials, but in the massless limit and without the external sources, there remains only one:
\begin{equation}\label{eq:LA6}
   F^2 {\cal L}^A_6 = \tilde C^{(6)}_1 \epsilon^{\mu\nu\alpha\beta}
    \langle  h_{\gamma\mu}[u^\gamma,\,u_\nu u_\alpha u_\beta] \rangle\,.
\end{equation}
And this is where the formal expansion in the odd-intrinsic sector ends in the literature. 

\subsection{The Singlet part}
As already mentioned, one of our tasks is to calculate the cyclically ordered amplitudes of the $SU(N)\times SU(N)/SU(N)$ chiral theory. At this point, we have summarized all known single trace monomials, and it seems straightforward gluing them together using the Feynman rules to get any amplitude with wanted order and multiplicity. There is, however, one catch. Gluing the vertices using the scalar propagators $i\delta^{ab}/p^2$ effectively means employing the completeness relations of $SU(N)$:
\begin{equation}\label{eq:complr}
    \sum_{a=1}^{N^2-1}\langle X t^a\rangle \langle t^a Y\rangle  = \langle XY\rangle - \frac1N \langle X\rangle \langle Y\rangle\,.
\end{equation}
We see that even using the single-trace ingredients, the resulting amplitude already at the tree-level may include multiple traces. This is a complication for us because, as we will see, we want to express our results using the stripped amplitudes, i.e., we want to separate the flavor structure (using one single trace of generators) and the kinematical structure (the stripped amplitude). There are several possible solutions, and their selection depends on the application. The simplest solution is to ignore this problem. It merely means that our stripped amplitude after dressing up with the single flavor trace will not correspond to the full amplitude but rather its single-flavor-trace part. For our purposes, this will be enough. In fact, for us the equivalent solution is to enlarge $SU(N)$ to $U(N)$, which means using unitarized $\Phi^U$ in~(\ref{eq:udef}):
\begin{equation}\label{eq:PhiU}
    \Phi(\phi)\quad \to\quad \Phi^U (\phi) = t^a\phi^a,\quad a=0,\ldots,N^2-1\,,
\end{equation}
with $t^0 = I_N/\sqrt{N}$. Now the completeness relation for $U(N)$ has only the first term with the single trace in~(\ref{eq:complr}). After all, physical motivation behind keeping the single trace monomials is due to large $N_C$ limit. And exactly in this limit the singlet $\phi^0$ (e.g. called $\eta'$ in $N=3$) becomes massless and can be naturally included in our multiplet of Goldstone particles.

On the other hand, if we are lucky, the second term in~(\ref{eq:complr}) cancels out for some amplitudes. This happens, for example, for all amplitudes at $O(p^2)$~\cite{Kampf:2013vha}. It means that there is no difference between calculating at this order amplitudes using $SU(N)$ or $U(N)$ -- all amplitudes with at least one external $\phi^0$ are zero. It is interesting to find out that the same also happens for the WZW Lagrangian~(\ref{eq:L5}). Working with the five-dimensional integrand, we can notice that $\Sigma_i^U$ with unitarized $\Phi^U$ (\ref{eq:PhiU}) can be expressed as
\begin{equation}
    ({\cal U}^{\scriptscriptstyle U})^\dagger \partial_i {\cal U}^{\scriptscriptstyle U} =  \frac{i}{F}\sqrt{\frac2N} \partial_i(z \phi^0) + {\cal U}^\dagger\partial_i {\cal U}\,.
\end{equation}
Contracting such five terms with the five-dimensional $\epsilon^{ijklm}$, we can confirm that terms with at least one $\phi^0$ cancel out after performing all permutations.

To summarize: in order to have at least one singlet in a vertex, one needs to go to at least $O(p^4)$ order in the even sector and $O(p^6)$ in the odd. Concerning the amplitudes with $\phi^0$ hidden in the internal lines -- we would need at least two such vertices. So our stripped amplitudes can be used for the full flavor-combination amplitudes up to and including $O(p^4)$ and $O(p^6)$ for the even and odd sector, respectively. In order to obtain a full multi-trace amplitude also beyond these orders, we refer to methods developed and used in~\cite{Bijnens:2019eze}.

\section{Bottom-up: basis}
In the previous section, we have summarized Lagrangian as known in both sectors of ChPT. We can now proceed to the textbook recipes and create off-shell vertices and then compose Feynman diagrams. We can thus obtain the tree-level amplitudes. The way offered here is different, though the objective is the same: the amplitudes. First, we will start with a replacement for the off-shell vertices. A similar strategy was already taken in literature, namely for the 4pt monomials up to $O(p^4)$ studied in \cite{Low:2019ynd}.

\subsection{Basis for kinematical variables}\label{sec:basisforkin}
The lowest possible point amplitudes are the 4-pt vertices (no 3-pt due to $p_1\cdot p_2\sim (p_1+p_2)^2=p_3^2 =0$, etc.). Note also that due to the parity conservation all even-pt vertices will be built of $(p_i\cdot p_j)$ products and the odd vertices must be proportional to one Levi-Civita tensor $\epsilon^{\mu\nu\alpha\beta}$. Starting with the 4pt $O(p^2)$ we will summarize briefly the construction of all $n$-pt $O(p^m)$ vertices up to $O(p^8)$, i.e. corresponding to monomials summarized in Tab.~\ref{tab:numberofpions1}. Note that we will use mainly the following convention for the kinematical variables:
\begin{equation}
    s_{ij} = (p_i + p_j)^2\,,
\end{equation}
which has a possible extension to $s_{ijk\ldots} = (p_i+p_j+p_k +\ldots)^2$.

The algorithm to construct the basis is quite simple. At the given order for the $n$-pt vertex, say at $O(p^2)$ for 4pt, we create an appropriate monomial and sum it over the cyclic momenta. For example $s_{12}$ gives us: $s_{12}+s_{23}+s_{34}+s_{41} = 2 (s_{12}+s_{23})$. If it is linearly independent from the existing basis, it is added to the basis. We scan over all possible combinations of monomials. What is, however, important is that we check the independence using the on-shell 4-dim kinematics, which means for example that $p_i^2=0$. The above-mentioned basis thus contains only one term. Another possible combination $s_{13}$ is not linearly independent due to the on-shell relation $s_{12}+s_{13}+s_{23}=0$. Note that our restriction to $D=4$ means that we automatically incorporate the non-linear constraints due to Gram determinants for higher point amplitudes. If we were not constrained by dimensions, we would have for the given $n$-point scattering
$
\frac12 (n-3) n \text{ terms}\,,
$
while for $D=4$ there are $3n-10$ terms. A difference starts at 6pt and is due to the Gram determinant relation
$$
\det{s_{ij}}|_{i,j=1\ldots6} = 0\,,
$$
which gives one constraint among nine otherwise independent $s_{ij}$, chosen for example as $s_{12},s_{13},s_{14},s_{15},s_{23},s_{24},s_{25},s_{34};s_{35}$. And similarly for higher-point kinematics.

The number of vertices is given in the following table~\ref{tab:numberofsijtermseven}. 
\begin{table}[tbh]
\begin{center}  
\begin{tabular}{|c|c|c|c|}
\hline
   &$n-$point & \# all terms & \# independent \\ \hline 
$p^2$ & 4 & 3 & 1  \\
\hline
$p^4$ & 4 & 6 & 2  \\
\hline
$p^6$ & 4 & 10 & 2   \\
      & 6 & 220 & 22 \\
\hline
$p^8$ & 4 & 15 & 3 \\
      & 6 & 715 & 58 \\
      & 8 & 10626 & 621\\
\hline
    \end{tabular}
    \caption{Number of possible monomials composed of products of $s_{ij}$. The all-terms column summarizes all combinations of the products at the given order, whereas in the last column the on-shellness, cyclicity, Hermiticity and the 4-dim restrictions are imposed.}\label{tab:numberofsijtermseven}
  \end{center}
\end{table}
There we have summarized the number of all possible terms unconstrained by on-shell relations (though we do not consider parameters $s_{ii}$) and 4-dim kinematics. For example, in the first line, 3 stands for $s_{12},s_{13},s_{23}$ schematically, in the second line 6 is $s_{12} s_{12},s_{12} s_{13},s_{12} s_{23},s_{13} s_{13},s_{13} s_{23},s_{23} s_{23}$, and similarly for other lines. Now we continue with the algorithm mentioned above: take the given term and sum cyclically over momenta, as the form of the stripped vertices dictates this property (see definition in~(\ref{eq:vertexordered})). Another important property is the dihedral reflection invariance or equivalently the invariance under the reverse ordering. This follows from the Hermiticity of Lagrangian and corresponding full vertices: conjugating them in~(\ref{eq:vertexordered}) leads to the reverse ordering of the stripped vertices (due to the trace of the Hermitian generators $t^a$)\footnote{We have tacitly assumed that all constants are real. If some of them have an imaginary part, we would have an extra minus sign for the conjugate in order to keep the Lagrangian Hermitian. Such terms would violate a charge conjugation (n.b. $\Phi^C\to\Phi^T$), but they are still parity invariant by construction. We will briefly return to this possibility in summarizing the number of terms in the following section and denote those terms as CP-odd.}.
So the final step of the algorithm is adding the Hermitian conjugate (i.e., reverse ordering) and at last checking the independence of the basis. The number of terms obtained is summarized in the last column. 

Here we will also cover the odd-intrinsic parity sector, see Tab.~\ref{tab:numberofsijtermsodd}.
\begin{table}[tbh]
\begin{center}  
\begin{tabular}{|c|c|c|c|}
\hline
   &$n-$point & \# all terms & \# independent \\ \hline 
$p^4$ & 5 & 1 & 1  \\
\hline
$p^6$ & 5 & 6 & 1   \\
\hline
$p^8$ & 5 & 21 & 3 \\
      & 7 & 1800 & 90 \\
\hline
    \end{tabular}
    \caption{Same as the previous table -- now for monomials composed of cyclic sum of  products of a Levi-Civita tensor and $s_{ij}$ for odd-number of pions.}\label{tab:numberofsijtermsodd}
  \end{center}
\end{table}
We have followed exactly same steps as in the previous even case, with only one difference, the insertion of the Levi-Civita tensor.

Of course we are still not done, those monomials represent only our ``raw data''. We will carve them out in the following text using the condition of the Adler zero.

\subsection{Amplitudes: general discussion}
Comparing the last columns of Tab.~\ref{tab:numberofpions1} and Tab.~\ref{tab:numberofsijtermseven} we see that the basis constructed in the previous section is an overdetermined system from the point of view of Goldstone boson vertices. We will discuss here how to approach the goal of creating combinations of the vertices that would comply with the dynamics of the Goldstone bosons. Note, however, that the 4pt numbers agree already without doing anything. This is connected with the fact that the 4pt kinematics is very restrictive. There are only two variables, conventionally $s=s_{12}$ and $t=s_{13}$ and it seems that for any combination at any order $s^i t^j$ can be connected (and this connection is one-to-one) with monomials in the Lagrangian of the form
\begin{equation}
    \sum_{\text{all possible contractions}}\langle h_{\mu\ldots}h_{\nu\ldots}h_{\rho\ldots}h_{\sigma\ldots}\rangle\,,
\end{equation}
where $h_\mu$ with only one Lorentz index is simply $u_\mu$ and $h_{\mu\ldots}$ with higher number of indices is defined as in~(\ref{eq:hmnr})
\begin{equation}\label{eq:hmnr2}
    h_{\mu\ldots\sigma\rho} = \nabla_\mu\ldots \nabla_\sigma u_\rho + \text{sym} \,.
\end{equation}
This is stated as a conjecture, but it can be understood also from the amplitude point of view as we will discuss below.

This brings us to a crucial point of this article. We have to demand certain conditions in order to comply with the chiral symmetry or, in other words, to limit the kinematical terms represented by Tab.~\ref{tab:numberofsijtermseven} to terms coming from the chiral Lagrangian, as summarized in Tab.~\ref{tab:numberofpions1}. We want to work directly with something ``physical'' using the on-shell data. To do so, we will be gluing together the on-shell vertices (our raw data) and getting amplitudes that would represent our physical data once they comply with certain conditions. It can be shown \cite{Kampf:2012fn, Kampf:2013vha, Cheung:2014dqa, Bijnens:2019eze} that the demanded condition has to be the Adler zero \cite{Adler:1969er}. For the $n$-point amplitudes with $n\geq6$ we can generalize it. Instead of studying one single soft limit of one external leg, we will be studying simultaneously single-soft limits of all $n$-legs at once. We will use the following trick, first introduced in \cite{Cheung:2015ota}. For $n>D+1$ it is possible to define a ``rescaling shift'' on all external lines:
\begin{equation}
    p_i \to p_i (1-z a_i)\,,
\end{equation}
with distinct $a_i$. There are always $n-D$ parameters we can set, and the rest is fixed by the condition
\begin{equation}
    \sum_{i=1}^n a_i p_i =0\,,
\end{equation}
which has to be valid in order to maintain the momentum conservation. As we want to study a {\it single\/} soft limit we have to ensure that all $a_i$ are distinct. For $D=4$ and a general kinematic configuration this is always possible starting with the 6-point amplitudes. Then the Adler zero can be formulated as a condition
\begin{equation}
    A_{n\geq6}(z) \sim (1-z a_i)\,, \quad \text{with }z\to1/a_i \,,
\end{equation}
valid for all $a_i$, $i=1,\ldots,n$. The amplitudes vanish in the first order in $z$ which brings us to study its analytical properties. As all $p_i$s are shifted and depend linearly on $z$, it is trivial to state that  shifted amplitudes behave at infinity as
\begin{equation}
    A_n(z) \to z^r\,, \quad \text{for }z\to\infty\,,
\end{equation}
where $r$ stands for the order, i.e. it behaves as $z^2$ for $O(p^2)$, as $z^4$ for $O(p^4)$, etc.
Now we can apply the Cauchy's theorem with an extra denominator factor to compensate the ``bad'' high-energy behavior without introducing extra poles:
\begin{equation}\label{eq:ct}
    \oint \frac{dz}{z}\frac{A_n(z)}{F_n(z)} = 0\,.
\end{equation}
The compensated factor is by construction given by
\begin{equation}
    F_n(z) = \prod_{i=1}^n (1-a_i z)\,.
\end{equation}
This can be further used in establishing the soft recursion using the soft behavior similar to the BCFW recursion relations based on the collinear shifts~\cite{Britto:2005fq}.
One can always continue, though, using the direct evaluation of Feynman diagrams. However, it is important methodologically -- it can easily explain why in preparing Tab.~\ref{tab:numberofsijtermseven} we have stopped at certain orders, without the necessity to discuss algebraic arguments in Lagrangian (i.e., using the form of $h_{\mu\ldots}$ symbols). The Cauchy integral formula tells us what input is and what has to be fixed. Starting with an order $O(p^2)$ and 6pt amplitudes. Then
$A_6\sim z^2$ and $F_6(z)\sim z^6$, so the 6pt amplitudes (and all higher orders as well) must be fully reconstructible from factorization channels only, i.e. from the $O(p^2)$ vertices/Lagrangian. Similarly, the $O(p^4)$ order, all amplitudes with $n\geq6$ are reconstructible from $O(p^2)$ and $O(p^4)$. For the order $O(p^6)$, the inputs are $O(p^2),O(p^4),O(p^6)$ vertices, but all amplitudes with $n\geq8$ are fully reconstructible. This concludes our amplitude explication of why we need to consider only those orders as summarized in Tab.~\ref{tab:numberofsijtermseven} and can be readily extended to all orders.

\section{Amplitudes}
We have summarized the basis of independent on-shell vertices based merely on the power-counting and without any connection to Lagrangian. 
Apart from work done in~\cite{Dai:2020cpk}, which also studies the operator basis using the bottom-up approach, we are in $D=4$ and are thus automatically incorporating the relations given by the Gram (sub)determinants.

Now we will focus on the stripped amplitudes. The conventional full amplitude for a given combination of flavors $a_1\ldots a_n$ at given order $O(p^r)$ is obtained by summing over all permutations modulo a cyclic permutation:
\begin{equation}\label{eq:fullA}
    A^{(r)a_1\ldots a_n}_n (p_1,\ldots,p_n)= \sum_{\sigma\in S_n/Z_n}
    A^{(r)}_n (p_{\sigma(1)},\ldots, p_{\sigma(n)})\langle t^{a_{\sigma(1)}}\ldots t^{a_{\sigma(n)}} \rangle \,.
\end{equation}
First, we will start with the even sector and, for the moment, will completely ignore the anomaly, which will be discussed later in two subsections.

\subsection{Even sector}

The simplest example is the 4pt amplitude at the LO $O(p^2)$. We can use the Cayley parametrization (cf.~(\ref{eq:V4})) or any other as the stripped amplitudes are ``physical quantities'' \cite{Kampf:2013vha, Bijnens:2019eze} -- meaning that they do not depend on parametrization:
\begin{equation}
    A^{(2)}_4 = -\frac{1}{2F^2} s_{13}\,.
\end{equation}
Note that this is a reduced or stripped amplitude with cyclically ordered momenta and we have to use~(\ref{eq:fullA}) if we need the conventional 4pt amplitude with given flavors.

\begin{figure}[htb]
\centering
\includegraphics[width=14cm]{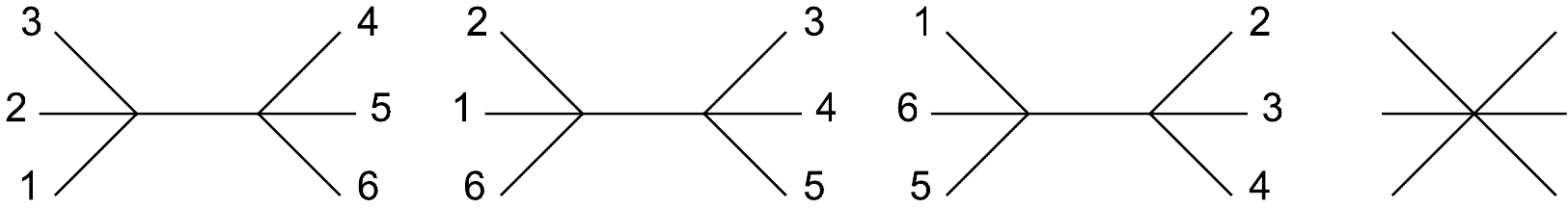}
\caption{The Feynman diagrams for the 6pt $O(p^2)$ scattering.}
\label{fig:feyndiag26}
\end{figure}

The simplest {\it non-trivial\/} example is the 6pt amplitude at $O(p^2)$. Using the standard Feynman diagram techniques, one has to calculate diagrams depicted in Fig.~\ref{fig:feyndiag26}.
Note that all vertices are governed by one constant $F$ corresponding to definitions in~(\ref{eq:L2}) and~(\ref{eq:udef}). 
The amplitude reads
\begin{equation}
    4 F^4 A_6^{(2)} = -\frac{s_{13}s_{46}}{s_{123}} -\frac{s_{26}s_{35}}{s_{345}}
    -\frac{s_{15}s_{24}}{s_{234}}
    +\frac12 (s_{13}+s_{15}+s_{24}+s_{26}+s_{35}+s_{46})\,.
\end{equation}
This result can be easily obtained using the standard Feynman rules or soft bootstrap (for details, see the original work \cite{Cheung:2015ota}).
We can now continue to higher orders. However, the purpose of this article is not to summarize the individual amplitudes (for the interested reader, we refer to \cite{Bijnens:2019eze}), but to reduce the basis, so it guarantees the Adler zero. For the lowest order, it was trivial, and the $O(p^4)$ is also trivial. We obtained two independent terms, and so has also the final basis (see Tab.~\ref{tab:numberofpions1}). However, we can ask if we can verify it independently. For this, we have employed the bonus relations. These relations can be applied in situations when we can drop $1/z$ in the Cauchy's theorem~(\ref{eq:ct}), and the recursion leads to relations among amplitudes with the same multiplicity~\cite{Cheung:2015ota}. At this order, it means to calculate $O(p^2)\times O(p^4)$ contribution at the 6pt level. We have checked that the two parameters are indeed independent.

\begin{figure}[htb]
\centering
\includegraphics[width=10cm]{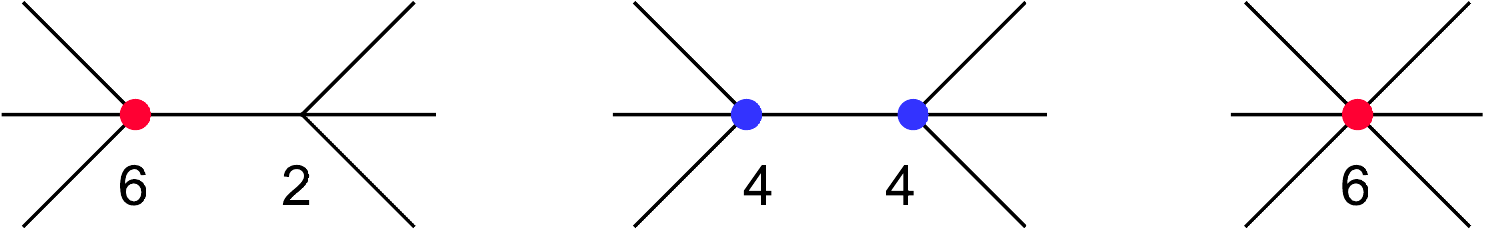}
\caption{Possible topologies of the Feynman diagrams for the 6pt $O(p^6)$ scattering. The numbers indicate the order of corresponding vertices.}
\label{fig:feyndiag66}
\end{figure}

The first nontrivial order is the NNLO, the $O(p^6)$. Indeed, comparing Tables~\ref{tab:numberofpions1} and~\ref{tab:numberofsijtermseven} 
we see that the first line to tame is the fourth line, i.e., the $O(p^6)$ order of the 6pt amplitude. To calculate this amplitude directly one has to evaluate the diagrams depicted schematically in Fig.~\ref{fig:feyndiag66}.
Note, however, similarly to the previous $O(p^2)$ calculation, we are calculating the stripped amplitudes, so we have to sum over all cyclic permutations. This means that the first diagram is, in fact, the sum of six diagrams, while the second topology sums over three diagrams only (with propagators (1+2+3), (6+1+2), (5+6+1)), because the two vertices are the same. The individual contributions can also be written schematically as
\begin{equation}\label{eq:A66}
    F^8 A_6^{(6)} =  \sum_{i=1}^2 C_i^{(6)} X_i^{(6)} + \sum_{i,j=1}^2 L_i L_j X_{ij}^{(6)} + \sum_{i=3}^{24} C_i^{(6)}\,,
\end{equation}
where every term corresponds precisely to the Feynman diagram topologies in Fig.~\ref{fig:feyndiag66}. The first two terms, factorization diagrams, are known, the last contact term contains 22 coefficients that must be fixed in order to comply with the Adler zero. We have 22 unknown constants to fix and they are connected with linearly independent monomials. Although we have put a momentum of one pion soft $p\to0$ and reduced the 6pt to 5pt kinematics, it does not mean it corresponds now to terms described in Tab.~\ref{tab:numberofsijtermsodd} (cf. 5-pt $O(p^6)$ order) as the cyclic symmetry is not automatically restored. This makes, on the other hand, the relation~(\ref{eq:A66}) so powerful. The easiest way how to solve it is to work numerically, i.e., generate at least 22 independent equations (similarly as done, e.g., in \cite{Cheung:2016drk}). This system of linear equations can then be easily reduced to the final set of independent $C_i^{(6)}$s. Our calculation reduced the number of 22 down to 5, and thus we have confirmed the ChPT construction. 
Let us note that~(\ref{eq:A66}) represents the full 6pt $O(p^6)$ amplitude. However, for our purpose -- i.e. to reduce the base of the $O(p^6)$ order, we do not need the $O(p^4)$-vertices insertion as these are already in their minimal base form. In other words we can set $L_i=0$ in Fig.~\ref{fig:feyndiag66} or~(\ref{eq:A66}) without loss of generality. We can, and we will in the following, turn off everything that is already in its minimal form. The only exception is the $O(p^2)$ order (via the parameter $F$), which must be kept as it also includes the kinetic term.

\begin{figure}[htb]
\centering
\includegraphics[width=7cm]{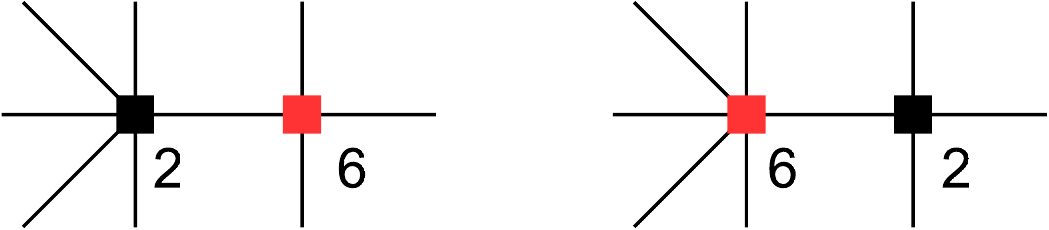}
\caption{Schematic representation of $A_L \times A_R$ in the bonus relations for the 8pt amplitude at $O(p^6)$ (with $L_i=0$). The box symbol indicates the corresponding full amplitude and the number its order.}
\label{fig:bonus6}
\end{figure}

Another possibility of how to check the result or how to verify that this 5 is a final number is via the bonus relations. Diagrammatically we have ``left'' and ``right'' amplitudes depicted in Fig.~\ref{fig:bonus6}.
Note that squares in this figure represent the amplitudes at $O(p^2)$ or $O(p^6)$ order, respectively (of course, the 4pt ``right-hand sides'' coincide with the vertices). We have verified that the connected bonus relations lead to no other supplement conditions on the $O(p^6)$ constants and are automatically satisfied once the amplitude $A_6^{(6)}$ in~(\ref{eq:A66}) comply with the Adler zero. Let us also stress that the bonus relations are satisfied once both topologies in Fig.~\ref{fig:bonus6} are combined, and the cyclic sum is performed. So it is another nontrivial test of our calculation. 
The order $O(p^6)$ seems thus fairly verified and can be trusted to give any amplitude also beyond the defining orders. The first such amplitude entirely fixed by the lower vertices is the 8pt one $A^{(6)}_8$ and can be obtained easily using the soft bootstrap. We have performed this calculation and will briefly comment on it also below in Sec.~\ref{sec:mixsector}.

\begin{figure}[htb]
\centering
\includegraphics[width=10cm]{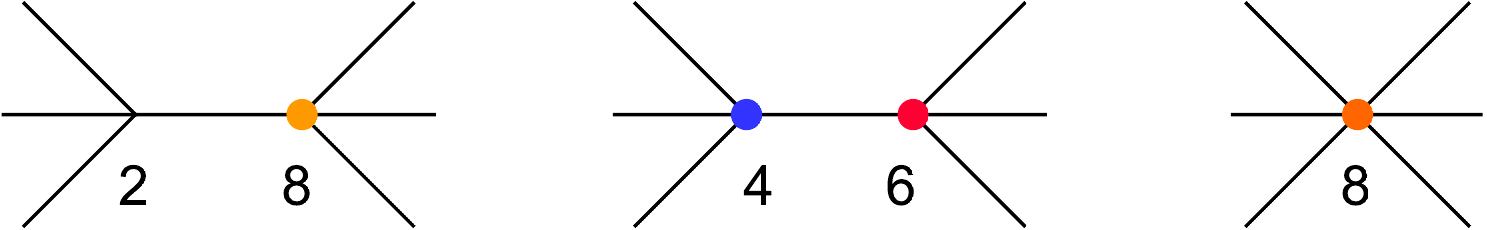}
\caption{Feynman diagrams for the 6pt $O(p^8)$ scattering. The numbers indicate the order of corresponding vertices.}
\label{fig:feyndiag86}
\end{figure}

Let us turn our attention to the last available order in the literature, NNNLO, i.e., $O(p^8)$. We will start with the full tree-level 6pt amplitude, depicted in Fig.~\ref{fig:feyndiag86}.
As above, we can disregard the contributions with $O(p^4)$ and $O(p^6)$ vertices (the second diagram). Demanding the Adler zero, we got constraints on the $O(p^8)$ constants $C_i^{(8)}$. The 4pt vertex parametrization stays as in the previous case without a change (i.e., parametrized by three constants), whereas the 6pt is reduced from 58 to 22. Note that this number is obtained if the Hermiticity was imposed. Explicitly, if the Hermiticity was not demanded, we would have 89 terms instead of 58 in Tab.~\ref{tab:numberofsijtermseven} and now we would get 29 terms, with extra 7 CP-odd terms. Interestingly, after demanding the Adler zero there are no CP-odd terms for lower orders.

\begin{figure}[htb]
\centering
\includegraphics[width=13cm]{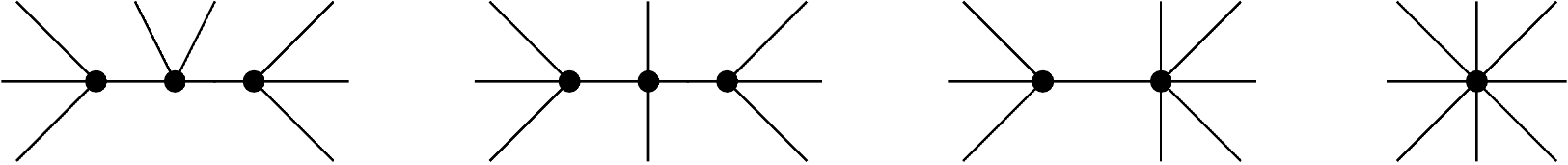}
\caption{Possible topologies of the Feynman diagrams for the 8pt $O(p^8)$ scattering. The black circle now represents the appropriate $O(p^2)$, $O(p^4)$, $O(p^6)$ or $O(p^8)$ vertex.}
\label{fig:feyndiag88}
\end{figure}

\begin{figure}[htb]
\centering
\includegraphics[width=11cm]{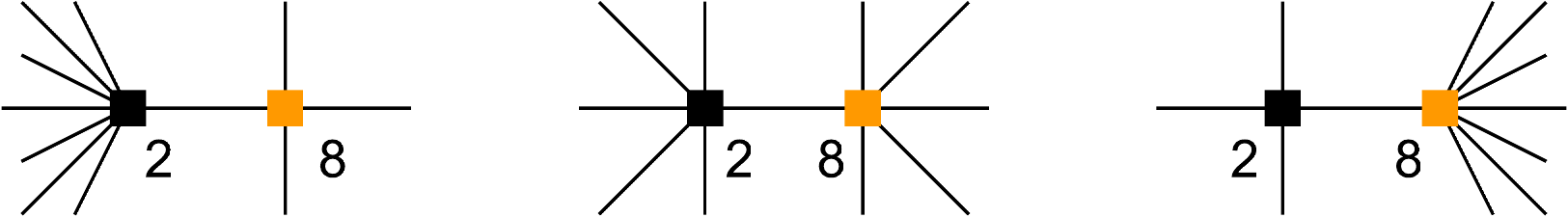}
\caption{Schematic representation of $A_L \times A_R$ in the bonus relations for the 10pt amplitude at $O(p^8)$. The numbers indicate the order of corresponding amplitudes.}
\label{fig:bonus8}
\end{figure}

The 8pt Feynman diagrams to calculate are depicted schematically in Fig.~\ref{fig:feyndiag88}.
The Adler zero condition leads to the considerable reduction from 621 to 17 monomials (from 1128 to 18, if the CP-odd terms allowed). Again as in the previous $O(p^6)$ case we can check our result employing the bonus relations. Technically it means to go to 10pt amplitudes and check the topologies schematically shown in Fig.~\ref{fig:bonus8}.
We performed this highly non-trivial check and verified that the basis is stable without any supplement constraint.

We summarize the situation in the even sector up to NNNLO by comparing the numbers of minimal basis obtained using our direct amplitude oriented construction and the existing ChPT Lagrangian in Table~\ref{tab:numberofpions2}.
\begin{table}[t]
\begin{center}  
\begin{tabular}{|c|c|c|c|}
\hline
   &\#mesons&\begin{tabular}{@{}c@{}}\#terms\\using amplitudes\end{tabular} &\begin{tabular}{@{}c@{}}\#terms\\using ChPT\end{tabular} \\ \hline 
$p^2$ & 4 & 1 & 1  \\
\hline
$p^4$ & 4 & 2 & 2   \\
\hline
$p^6$ & 4 & 2 & 2   \\
      & 6 & 5 & 5 \\
\hline
$p^8$ & 4 & 3 & 3  \\
      & 6 & 22(7) & 22 \\
      & 8 & 17(1) & 17 \\
\hline
    \end{tabular}
   \renewcommand{\arraystretch}{1}
     
    \caption{Number of monomials that produce vertices starting at the given number of mesons based on the tree-level Adler zero. The numbers in parentheses summarize extra terms needed if the Hermiticity condition was not applied.
    The last column is copied from Tab.~\ref{tab:numberofpions1} summarizes the situation obtained from ChPT literature as discussed in Sec.~\ref{sec:even}.}\label{tab:numberofpions2}
  \end{center}
\end{table}
%

\subsubsection*{$O(p^{10})$ order}
As we can see, our construction works perfectly up to $O(p^8)$ and also agrees with a similar bottom-up approach in \cite{Dai:2020cpk}. However, we can ask why we agree so well if our procedure is unique in the sense that we work in $D=4$, but both the ChPT-top-down construction and the one in~\cite{Dai:2020cpk} is valid only for $D\geq n-1$. Shouldn't we see some discrepancy starting with multiplicity $n=6$ already at $O(p^6)$ or at least at $O(p^8)$? The answer is apparently no. The reason is that the relations due to zeros of the Gram sub-determinants (i.e., highly non-linear relations for the $s_{ij}$) cannot be directly applied to the {\it linear\/} relations among monomials. Our procedure represents clear proof of this statement. There is, however, one loophole in the previous argument: any Gram (sub)determinant produces a linear relation for the products of $s_{ij}$ and if this matches the studied order, we have an extra condition on monomials owing to $D=4$. As the lowest-order of relations due to the Gram determinants is $s_{ij}^5$, the lowest order to see some discrepancy is thus $O(p^{10})$. 

It is apparent that the $O(p^{10})$ is in some sense important as it brings a new effect in the basis construction. It is thus essential to see and verify how this works explicitly. It justifies the need to push our algorithm and the use of the computer time for this order. It is also enabled by an existing study in~\cite{Dai:2020cpk} which performed this analysis without dimensional constraints. The numbers we have obtained are summarized in Tab.~\ref{tab:order10}.
\begin{table}[t]
\begin{center}  
\begin{tabular}{|c|c|c|c|}
\hline
   &\#mesons&\begin{tabular}{@{}c@{}}\#terms for \\ $D=4$\end{tabular} &\begin{tabular}{@{}c@{}}\#terms for\\ $D\geq 5$ \end{tabular} \\ \hline 
$p^{10}$ & 4 & 3 & 3  \\
      & 6 & 70 & 71 \\
      & 8 & 248 & 255 \\
      & 10 & 73 & 79 \\
\hline
    \end{tabular}
   \renewcommand{\arraystretch}{1}
     
    \caption{Number of monomials that produce vertices starting at the given number of mesons based on the tree-level Adler zero. Note the effect of the dimension. The $D=4$ column summarizes the results of our algorithm and
    the last column, representing monomials unconstrained by dimension, is obtained from \cite{Dai:2020cpk}.}\label{tab:order10}
  \end{center}
\end{table}
As expected, we can see that the 4pt case is without a change (see next subsection for all orders). In the 6pt case, we can see finally a difference given by one term -- exactly as expected due to one linear $O(p^{10})$ Gram relation. For higher multiplicities: we have a difference of 7 and 6 relations for $n=8$ and $n=10$, respectively. We have explicitly verified that these relations are again due to the independent sets of the Gram subdeterminants. It would be interesting to see the explicit construction of the sets as it reminds the terms obtained using the techniques of the multi-Galileon theories (cf. for example \cite{Kampf:2020tne}).

\subsubsection*{4pt to all orders}
As the 4pt-order amplitude is the contact vertex, it seems simple to try to prescribe the amplitude structure to all orders. Indeed,   at the $O(p^r)$ order, the cyclic amplitude is given by (it can be easily proved by expressing it using two independent variables and employing the symmetry under cyclicity)
\begin{equation}\label{eq:A4all}
    A_4^{(r)} (s,t,u)  = \sum_{i=0}^{\lfloor r/4 \rfloor} \frac{c_{i+1}^{(r)}}{F^r} (s^{r/2-i} + u^{r/2-i})t^i \,,
\end{equation}
where $\lfloor..\rfloor$ represents the integer part. We have used a standard Mandelstam definition for the 4pt scattering $s=s_{12}$, $u=s_{14}$ and $t=s_{13}$. Of course $s+t+u=0$ and the cyclicity interchanges $(s,t,u)\leftrightarrow(u,t,s)$. We had to introduce a new symbol $c^{(r)}_i$ in order not to mix up with already defined Lagrangians in Sec.~\ref{sec:even}. However, it is easy to work out the relations:
\begin{equation}\label{eq:cC}
    c_1^{(2)} = 1/2,\;
    c_1^{(4)} = 2 L_3,\;
    c_2^{(4)} = -4 L_0,\;
    c_1^{(6)} = -16 C_2^{(6)},\;
    c_2^{(6)} = -4 C_1^{(6)}-24 C_2^{(6)}.
\end{equation}

Another interesting result hidden in the general formula is the number of terms:
\begin{equation}
    O(p^r):\qquad \text{\# term} =  \lfloor r/4 \rfloor + 1 \,,
\end{equation}
equal to $1,2,2,3,3,4,4,\ldots$ in agreement with the direct calculation as summarized in Tab.~\ref{tab:numberofpions2} and Tab.~\ref{tab:order10}.

\subsubsection{Amplitudes in a special kinematical point}\label{sec:speckinpoin}
On the way to cut the parametric space of higher orders, we have calculated many amplitudes but did not show them explicitly here. They can be obtained using the automatic program\footnote{They can also be obtained by contacting the author of this paper.} introduced in \cite{Bijnens:2019eze}. Still, it might be helpful to see the structure of the amplitudes more explicitly than given only schematically as e.g. for the 6pt in~(\ref{eq:A66}). For this, we will introduce a special kinematical point at which we express our amplitudes. There are certainly many choices for this point. We found it useful to pick a so-called democratic point where all relevant kinematical parameters are equal. In some sense, it should generalize the center of the Dalitz plot for the 4pt amplitudes, however, the naive one $s=t=u$ is not very useful for us. Instead, we use a geometrical structure, a kinematic polytope where the face is connected with a planar variable $X_{ij}$ \cite{Arkani-Hamed:2017mur}:
\begin{equation}\label{eq:Xij}
    X_{ij}  \equiv s_{i(i+1)\ldots (j-1)} = (p_i + p_{i+1} + \ldots + p_{j-1} )^2\,.
\end{equation}
Demanding a regular polytope, i.e. demanding that all non-zero $X_{ij}$ should be equal, we can reduce the number of independent $s_{ij}$ to one parameter. E.g. for $n=4$ we get
\begin{equation}
    \bar s\equiv s_{12} = -\frac12 s_{13} = s_{14}\,,
\end{equation}
while for $n=6$:
\begin{equation}
    \bar s\equiv s_{12} = - s_{13} = - s_{15} = s_{23} = - s_{24} = s_{34} = - s_{35} = s_{45},\quad s_{14}=s_{25}=0\,.
\end{equation}
One can easily calculate the higher multiplicities. It is interesting that for $n=6$, the corresponding {\it nona\/}hedron stays regular also in $D=4$ where we have to fulfill one extra constrain due to the Gram determinant. However, for higher $n$ this is not possible, and the regular polytopes are always squeezed when moving to $D=4$. The amplitudes are
\\
for $n=4$:
\begin{align}
    &F^2 A^{(2)}_4 = \bar s\notag\\
    &F^4 A^{(4)}_4 = 4 \bar s^2 (4 L_0 + L_3)\notag\\
    &F^6 A^{(6)}_4 = 16 \bar s^3 (C^{(6)}_1 + 4 C^{(6)}_2)\,,
\end{align}
for $n=6$:
\begin{align}
    &F^4 A^{(2)}_6 = -\frac32 \bar s\notag\\
    &F^6 A^{(4)}_6 = -6 \bar s^2 (6 L_0 + L_3)\notag\\
    &F^8 A^{(6)}_6 = -6 \bar s^3 (4 C^{(6)}_1 + 20 C^{(6)}_2 + C^{(6)}_3 + C^{(6)}_4 + 32 L_0^2)\,,
\end{align}
and for $n=8$:
\begin{align}
    &F^6 A^{(2)}_8 = \frac52 \bar s\notag\\
    &F^8 A^{(4)}_8 = 10 \bar s^2 (8 L_0 + L_3)\notag\\
    &F^{10} A^{(6)}_8 = 4 \bar s^3 (12 C^{(6)}_1 + 64 C^{(6)}_2 + 3 C^{(6)}_3 + 4 C^{(6)}_4 + 192 L_0^2)\,.
\end{align}
A rule of thumb for actual values of the low-energy constants based on resonance saturation and phenomenology is $10^{-3}$ and $10^{-5}$ for $L_i$s and $C^{(6)}_i$s, respectively. We can see that, for example, the contribution of diagrams with two $O(p^4)$ insertions ($\sim L_0^2$) is of the same importance as contributions of diagrams with $C^{(6)}_i$ and none of the contributions can be easily omitted (without more precise knowledge of the individual values). 
\subsection{Odd sector}

In the odd sector, we have to follow the same techniques to check amplitudes at the given order and multiplicity, though there are small differences. First of all, the lowest valency is now the 5pt vertex instead of the 4pt of the even sector. It seems it can be checked nontrivially for the Adler zero and it can give us some constraints. However, we can easily show that it is again trivial. The reason is that the 5pt vertex (= amplitude) is due to parity conservation always ``saturated'' by one Levi-Civita and thus automatically zero for any momenta going to zero. Its form can be written schematically as
\begin{equation}\label{eq:genA5}
    A^{(r)}_5 (1,2,3,4,5) = \epsilon^{1234} \sum_{ijkl} \tilde c_{ijkl}^{(r)} s_1^i s_2^j s_3^k s_4^l s_5^{r/2-i-j-k-l-2}\,,
\end{equation}
where $\epsilon^{1234} \equiv \epsilon^{p_1p_2p_3p_4} \equiv \epsilon^{\mu\nu\alpha\beta}p_{1\mu}p_{2\nu}p_{3\alpha}p_{4\beta}$ and $s_1,\ldots, s_5$ represent five independent scalar products of the 5-pt scattering. Number of such terms is thus proportional to combinations of $s_{ij}$ products corresponding to the order $r$ lower by four (due to four momenta in the Levi-Civita epsilon). Moreover, we must ensure that the Hermiticity is correctly incorporated -- the amplitude must be invariant by the reverse permutation. It can be easily calculated to any order and is given by (with the CP-odd terms in parentheses):  1,1,3,5(2),10(4),16(10),26(16),\ldots\footnote{This series of numbers has a nice geometrical interpretation: combinations of the turnover necklaces with 5 white stones and $n-1$ black stones.}.
We see that the beginning agrees with the explicitly calculated three lowest orders (cf. Tab.~\ref{tab:numberofsijtermsodd}).

As the anomalous sector is not yet well covered by literature, we will give more examples of amplitudes for lower orders and multiplicities.
The lowest order is fully determined by the chiral anomaly (cf. Appendix~\ref{sec:cayley})
\begin{equation}\label{eq:A45}
    A^{(4)}_5 =V_5(p_1,p_2,p_3,p_4)= \frac{N_C}{6\sqrt2\pi^2 F^5} \epsilon^{1234} \,.
\end{equation}
Higher orders depend on the convention of constants $\tilde c$ in~(\ref{eq:genA5}). For NLO this can be rooted back to Lagrangian~(\ref{eq:LA6}) and we can obtain
\begin{equation}
    A^{(6)}_5 = -8\sqrt2 \frac{\tilde C^{(6)}_1}{F^7}\epsilon^{1234} (s_{12}+s_{23}+s_{34}+s_{45}+s_{15})\,.
\end{equation}
For higher orders the canonical Lagrangian is not yet set. From Sec.~\ref{sec:basisforkin} we know that for $O(p^8)$ there are three constants, and we can for example take:
\begin{equation}
    A^{(8)}_5 = \epsilon^{1234} \bigl(\tilde c^{(8)}_1 s_{12}^2
    + \tilde c^{(8)}_2 s_{12}s_{23}
    + \tilde c^{(8)}_3 s_{12}s_{34} + \text{cycl} \bigr)
\end{equation}
and similarly for higher orders as already anticipated in~(\ref{eq:genA5}).

\begin{figure}[htb]
\centering
\includegraphics[width=6.5cm]{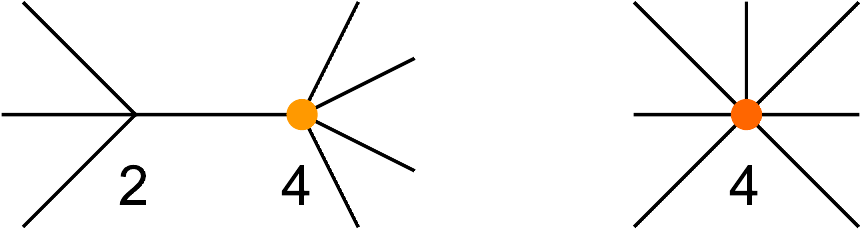}
\caption{Feynman diagrams for the 7pt $O(p^4)$ scattering. The numbers indicate the order of corresponding vertices and different color of odd number vertices is to accent the anomalous vertices.}
\label{fig:feyndiag47}
\end{figure}

Next interesting amplitude is naturally the $O(p^4)$ 7pt scattering. It can be obtained easily calculating following Feynman diagrams (Fig.~\ref{fig:feyndiag47})
using existing vertices, i.e.~(\ref{eq:V4}), (\ref{eq:V5}) and (\ref{eq:V7}), we can thus write
\begin{equation}
A^{(4)}_7 = \Bigl[ -\frac{N_C}{6\sqrt2 \pi^2 F^5} \epsilon^{1234}\frac{1}{(p_5+p_6+p_7)^2}\frac{-1}{2F^2} s_{57} + \text{cycl} \Bigr] + V_7  \,.
\end{equation}
One can verify that this amplitude fulfils the Adler zero. We have also verified that it can be obtained by the bottom-up approach, based only on the soft recursion, so it works similarly as in the even sector.

Nevertheless, this brings us to another important difference from the even sector. It is connected with shifted orders vs. multiplicities. There is one consequence that apart from the soft recursion, the bonus relations cannot be used so simply as in the even sector. For example for the NLO order $O(p^6)$, i.e. $r=6$, we can use the BCFW to calculate $n=7$ amplitude, but this multiplicity is not enough to get the bonus relation. For this, we would have to study at least $n=9$ (for order $r=6$).

\begin{figure}[htb]
\centering
\includegraphics[width=6.5cm]{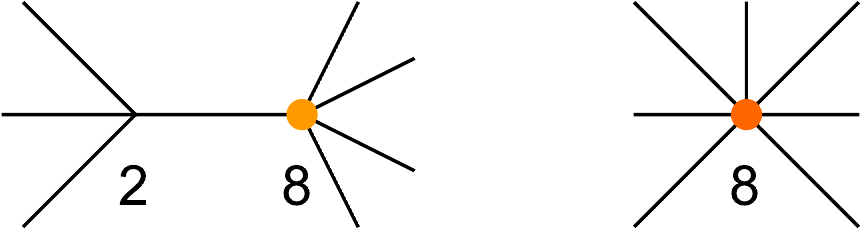}
\caption{Feynman diagrams for the 7pt $O(p^8)$ scattering with $\tilde C^{(8)}$ insertion. The numbers indicate the order of corresponding vertices.}
\label{fig:feyndiag87}
\end{figure}

From the bottom-up perspective, which is our main task here, the first real non-trivial order is thus NNLO, i.e. $O(p^8)$ -- it can bring something new. There the simplest non-trivial amplitude is the 7pt scattering, schematically expressed as:
\begin{equation}
    F^{11} A^{(8)}_7 = \sum_{i=1}^3 \tilde C^{(8)}_i \tilde X^{(8)}_i + \sum_{i=4}^{93} \tilde C^{(8)}_i
    + \sum (L_i \tilde C^{(6)}_j +L_i L_j \tilde C^{(4)}_k + C^{(4)}_i \tilde C^{(4)}_j)  \tilde X^{(8)}_{ij(k)}\,.
\end{equation}
If we switch off the known orders (i.e. the last sum) we have to calculate only diagrams depicted in Fig.~\ref{fig:feyndiag87}.
Demanding now the Adler zero, we obtain 75 non-trivial relations among $\tilde C^{(8)}_i$ constants, meaning that there are 3+15 independent constants. It is possible to calculate a 9pt $O(p^8)$ amplitude, however, it cannot give us any new constraints based on the above discussion of the bonus relations here. We would have to calculate 11pt scattering, which is for the moment beyond our possibilities. 
The present situation in the odd sector is summarized in Tab.~\ref{tab:numberofpions3}.

\begin{table}[t]
\begin{center}  
\begin{tabular}{|c|c|c|c|}
\hline
   &\#mesons&\begin{tabular}{@{}c@{}}\#terms\\using amplitudes\end{tabular} &\begin{tabular}{@{}c@{}}\#terms\\using ChPT\end{tabular} \\ \hline 
$p^4$ & 5 & 1 & 1   \\
\hline
$p^6$ & 5 & 1 & 1   \\
\hline
$p^8$ & 5 & 3 & ?  \\
      & 7 & 15(8) & ? \\
\hline
    \end{tabular}
   \renewcommand{\arraystretch}{1}
     
    \caption{Number of monomials that produce vertices starting at the given number of mesons based on the tree-level Adler zero. The number in parentheses counts the non-Hermitian monomials.
    The last column is copied from Tab.~\ref{tab:numberofpions1} summarizes the present situation of the ChPT literature as discussed in Sec.~\ref{sec:odd}.}\label{tab:numberofpions3}
  \end{center}
\end{table}
%
\subsection{Mixed sector}\label{sec:mixsector}
\begin{figure}[htb]
\centering
\includegraphics[width=3cm]{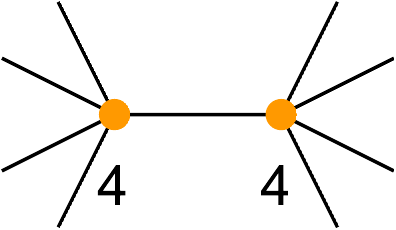}
\caption{Feynman diagrams for the 8pt $O(p^6)$ scattering with a double-insertion of the anomalous vertex $V_5$. The numbers indicate the order of this vertex.}
\label{fig:feyndiagAA}
\end{figure}

Of course, we can have only an even or odd number of external legs, so there cannot be a real mixed sector in multiplicity. Nevertheless, starting at $O(p^6)$ and at 8pt, the anomaly also enters the even sector. It means that to the result calculated using only even-sector vertices (for example, using the soft bootstrap as partially indicated in Fig.~\ref{fig:bonus6}), we have to add one diagram depicted in Fig.~\ref{fig:feyndiagAA}.
The amplitude is very simple and can be readily written as
\begin{equation}\label{eq:A86aa}
    A_8^{(6)A} = - V_5(1,2,3,4)\frac{1}{(p_1+p_2+p_3+p_4)^2} V_5(5,6,7,8) + \text{cycl}\,,
\end{equation}
where the anomalous vertex is given in~(\ref{eq:A45}). The complete amplitude at $O(p^6)$ is thus given as a sum of $A^{(6)}_8$  and $A_8^{(6)A}$.
It would be instructive to compare the anomalous contribution of the amplitude in comparison with the pure even sector. This is complicated by the fact that we are talking about 8pt scattering with 14 independent scalars $s_{ij}$ (if we are in $D=4$) or 20 (for general $D$). We will thus present this result only in one special kinematical point as introduced in Sec.~\ref{sec:speckinpoin}. We get
\begin{equation}
    A^{(6)\text{mix}}_8\bigl|_\text{spec.} =  \frac{\bar s^3}{F^{10}}\biggl( 48 C^{(6)}_1 + 256 C^{(6)}_2 + 12 C^{(6)}_3 + 16 C^{(6)}_4 + 768 L_0^2 - \frac{N_C^2}{9 \pi^4}  \biggr)\,.
\end{equation}
We see that the anomalous contribution is definitely not negligible, and using the rough estimate ($L_i\sim 10^{-3}$ and $C_i\sim 10^{-5}$), it seems dominant or at least of the same order.

\section{Further relations}

So far, the cyclicity was the fundamental simplification to study a general stripped amplitude (a function of $n$ momenta) as it enables to express $n!$ possible combinations of amplitudes from the $(n-1)!$ base. Schematically we can express relations between
$$
A_n ( \{p_1,\ldots,p_n\})\qquad \to \qquad A_n ( p_1,\{p_2\ldots,p_n\})\,,
$$
where $\{p_1,\ldots,p_n\}$ and $\{p_2,\ldots,p_n\}$ represent $n!$ and $(n-1)!$ permutations, respectively. From the top-down perspective, this is motivated and allowed by the large-$N_C$ arguments of QCD. At this point, it is just a coincidence that the same stripping down to the cyclically ordered amplitudes is also possible in Yang-Mills. However, we turn this around and ask if there are other relations valid in Yang-Mills or other theories that we can borrow and apply on our ChPT stripped amplitudes. 

\subsection{Kleiss-Kuijf and BCJ relations}
Indeed there exist relations well-known in other theories, worth applying also in ChPT. We can formulate it as a three-step procedure in reducing the base of $n!$ amplitudes $A_n(p_1,\ldots, p_n)$. If we take the cyclicity as a first step, then there are Kleiss-Kuijf (KK) relations \cite{Kleiss:1988ne} which take the base of amplitudes from $(n-1)!$ space down to $(n-2)!$ and finally Bern-Carrasco-Johansson (BCJ) relations \cite{Bern:2008qj} down to $(n-3)!$. 

The KK relations read (for simplicity dropping $p$, i.e. $p_i\to i$)
\begin{equation}\label{eq:KK}
    A_n(1,\{\alpha\},n,\{\beta\}) = (-1)^{|\beta|} \sum_{\text{OP } \{\alpha\}\cup \{\beta^T\}} A_n(1,\{\sigma\},n)\,,
\end{equation}
where $\{\alpha\}$ and $\{\beta\}$ are arbitrary sets of $2,\ldots,n-1$ momenta and $|\beta|$ represents length of the $\beta$ set. The ordered permutations (OP) of $\alpha$ and $\beta^T$ (reverse order of $\beta$ elements) keep ordering of $\alpha$ elements and {\it independently\/} of $\beta^T$ elements unchanged.

The BCJ relations are
\begin{equation}\label{eq:BCJ}
  s_{12} A_n(1,2,\ldots,n) +  \sum_{a=3}^{n-1} (s_{12}+s_{23}+\ldots+s_{2a}) A_n (1,3,\ldots,a,2,a+1,\ldots,n) = 0\,.
\end{equation}

\subsubsection*{4pt scattering}
The 4pt scattering is of special status as we know it up to all orders. It would be thus interesting to see how the parameter space changes in order to comply with constraints given by KK and BCJ relations. Starting with the KK, we will demand
\begin{equation}
    A_4^{(r)}(1,2,3,4)+A_4^{(r)}(1,2,4,3)+A_4^{(r)}(1,3,2,4) = 0\,.
\end{equation}
All other KK relations are either trivial or equivalent to this one. First of all the $O(p^2)$ order is automatically satisfied, proved for all multiplicity not only 4pt vertex. As we are focusing mainly on amplitudes up to $O(p^8)$ let us summarize the results of these amplitudes
\begin{align}
    \text{KK: }\; & F^4 A_4^{(4)} = c^{(4)}_2 \frac{ 2 s^2+2 s t-t^2}{2}\\
    & F^6 A_4^{(6)} = c^{(6)}_2 \frac{t (s^2+s t+t^2)}{2}\label{eq:KK64}\\
    & F^8 A_4^{(8)} = c^{(8)}_3 t^2 (s^2+u^2) +c^{(8)}_2 t (s^3+u^3) + (c^{(8)}_2-c^{(8)}_3) \frac{s^4+u^4}{2}\,. \label{eqA48kk}
\end{align}
However, there is no problem in calculating any order. The general form already visible at $O(p^8)$~(\ref{eqA48kk}) repeats to all orders:
\begin{equation}
    F^r A_4^{(r)\,\text{KK}} = \sum_{i+j=r/2} k_{ij} t^i (s^j + u^j)\,.
\end{equation}
It would be interesting to work out general formula for $k_{ij}$ coefficients to all orders and discuss its status.
The number of terms of such a theory is
\begin{equation}
    \text{KK: }\; O(p^r):\qquad \text{\# term} =  \lfloor (r-2)/6 \rfloor + 1 \,,
\end{equation}
or in a normal language: $1,1,1,2,2,2,3,3,3,4,4,4,\ldots$ (the first four agree with the amplitudes explicitly discussed above).

Let us proceed with the BCJ constraints. Again the $O(p^2)$ order is automatically satisfied (and the same for higher $n$-pt amplitudes). The amplitudes up to NNNLO:
\begin{align}
    \text{BCJ: }\; & F^4 A_4^{(4)} = 0\\
    & F^6 A_4^{(6)} = \frac{c^{(6)}_2}{4} t (s^2+t^2+u^2)\\
    & F^8 A_4^{(8)} = \frac{c^{(8)}_3}{3} t (s^3+t^3+u^3)\,,
\end{align}
or given by constraints on parameters $c^{(4)}_i$ (cf. also~(\ref{eq:cC})):
\begin{equation*}
    c^{(4)}_i = 0\,,\qquad
    2 c^{(6)}_1 = c^{(6)}_2 = -16 C^{(6)}_1\,,\qquad
    c^{(8)}_1 = 0,\; c^{(8)}_2 = c^{(8)}_3\,.
\end{equation*}
Notice the zero at the NLO order. It seems tempting to work out the higher orders as well and see some pattern. What is interesting is already the number of parameters at every order, given by the following (conjectured) formula
\begin{equation}
    \text{BCJ: }\; O(p^r):\qquad \text{\# term} =  \lfloor (r+2)/4 \rfloor - \lfloor (r+2)/6 \rfloor \,,
\end{equation}
equal for few orders to $1, 0, 1, 1, 1, 1, 2, 1, 2,\ldots$.
It is amusing that this formula coincides with the number of independent parameters of 4pt amplitudes lower by one order for a single scalar (with the full permutation). A quick check: the $O(p^4)$ order has 0 above. It should correspond to $O(p^2)$ fully permuted scalar amplitude which is $s+t+u$, indeed again zero. Our conjecture of its form is thus
\begin{equation}\label{A4rbcj}
    F^r A_4^{(r)\,\text{BCJ}} = \sum_{i=1}^{\lfloor (r+2)/4 \rfloor - \lfloor (r+2)/6 \rfloor} k^{(r)}_i t \,T^{(r-2)}(s,t,u)_i\,,
\end{equation}
where $T^{(r)}(s,t,u)_i$ represents the $i$-th monomial of the scalar 4pt amplitude. For curiosity: the first order with at least two monomials is $O(p^{12})$ and they read\footnote{The first proof of the form and number of these monomials, author is aware of, is due to J.Bijnens~\cite{bijnens:private}.}
\begin{equation}
    T^{(12)}(s,t,u) = (s^6+t^6+u^6\,,\; s^2 t^2 u^2)\,,
\end{equation}
(another candidates, e.g. $s^3 t^3 + s^3 u^3 + t^3 u^3$, are just linear combinations of the above two). It is natural to define the first monomials as
\begin{equation}
    T^{(r)}(s,t,u)_1 = s^{r/2} + t^{r/2} + u^{r/2} 
\end{equation}
and easily get the corresponding coefficients
\begin{align}
   \text{BCJ: }\qquad &k_1^{(2)} = -1/3\, c_1^{(2)}= -1/6,\quad
    k_1^{(4)} = \text{undef},\quad
    k_1^{(6)} = 1/4\, c_2^{(6)},\notag\\
    &k_1^{(8)} = 1/3\, c_3^{(8)},\quad
    k_1^{(10)} = -1/10\, c_3^{(10)},\quad
    k_1^{(12)} = -1/10\, c_4^{(12)},\notag\\
    &k_1^{(14)} = -3/28\, c_3^{(14)} + 2/7\,  c_4^{(14)},\quad
    k_2^{(14)} = 5/4\, c_3^{(14)} - 3\, c_4^{(14)}\,,\label{eq:kbcj}
\end{align}
which fully define the amplitudes $A_4^{(r)\,\text{BCJ}}$ for $r\le14$ in parameters of~(\ref{eq:A4all}). Of course no problem to calculate higher order coefficients, once the definition for $T^{(r-2)}(s,t,u)_i$ is provided. 

\subsubsection*{Higher multiplicities}
As for the $O(p^2)$ both KK and BCJ are satisfied for any number of multiplicities, we will study two higher orders: $O(p^4)$ and $O(p^6)$. 

For the $O(p^4)$ we have already solutions of KK or BCJ constraints from the previous subsection which we can express as
\begin{equation}\label{eq:KKcond4}
    \text{KK: }\; C^{(4)}_1 + C^{(4)}_2 = 0
\end{equation}
and in the case of BCJ both constants has to be zero:
\begin{equation}
    \text{BCJ: }\; C^{(4)}_1 = C^{(4)}_2 = 0\,.
\end{equation}
We may ask if a solution based on 4pt amplitudes is not violated for higher-pt amplitudes. For stronger BCJ this is trivially satisfied, but we can check KK relations~(\ref{eq:KK}) for higher $n$. We have indeed checked explicitly KK relations for $n=6$, i.e. for example
\begin{multline}\label{eq:KK6}
    A_6^{(4)}(1,2,3,4,5,6)
    +A_6^{(4)}(1,2,3,4,6,5)
    +A_6^{(4)}(1,2,3,5,4,6)\\
    +A_6^{(4)}(1,2,5,3,4,6)
    +A_6^{(4)}(1,5,2,3,4,6)=0
\end{multline}
and also for $n=8$ and they all confirmed~(\ref{eq:KKcond4}).

The first non-trivial order for higher multiplicity will be thus $O(p^6)$. Again the 4pt amplitude has been already discussed above, so let us turn to the 6pt case. We take the amplitude~(\ref{eq:A66}) complying with the Adler zero, the KK $O(p^4)$ constraint~(\ref{eq:KKcond4}) and the KK 4pt $O(p^6)$ condition
\begin{equation}
    \text{KK: }\; C^{(6)}_1 - 2 C^{(6)}_2 = 0\,,
\end{equation}
cf. also~(\ref{eq:KK64}). The amplitude now depends on one 4pt $O(p^4)$ constant (e.g. square of $C^{(4)}_1$), one 4pt $O(p^6)$ constant (e.g. $C^{(6)}_1$) and five 6pt constants $C^{(6)}_{3\ldots7}$. Plugging such amplitude to 6pt KK relations (cf.~\ref{eq:KK6}) we got following conditions on three 6pt constants
\begin{align}
    \text{KK: }\quad &C^{(6)}_5 = - C^{(6)}_1 - 3 C^{(6)}_3 - C^{(6)}_4\,,\quad
    C^{(6)}_6 = 2 C^{(6)}_1 + 6 C^{(6)}_3 + C^{(6)}_4\,, \notag\\
 &C^{(6)}_7 = - C^{(6)}_1 - 4 C^{(6)}_3 - C^{(6)}_4\,.
\end{align}
We have verified that the 8pt $O(p^6)$ KK relations hold too.

BCJ -- we will start directly with the 6pt $O(p^6)$ amplitude. In order to fulfill the BCJ relation~(\ref{eq:BCJ}) for $n=6$ we have to add to above constraints the following one
\begin{equation}
    C^{(6)}_3 = -\tfrac23 C^{(6)}_1\,,\quad C^{(6)}_4 = 0
\end{equation}
and again verified also at the 8pt (cf. Appendix~\ref{sec:ap8pt}). We see that the amplitude at this order depends only on one parameter. At the 6pt it reads
\begin{align}
   F^8 A^{(6)\text{BCJ}}_6 &= -4 C^{(6)}_1 \Bigl(
   \frac{s_{13}(s_{12}^2 + s_{12} s_{13} +  s_{13}^2) s_{46}}{s_{123}} 
   +s_{13}^2 s_{45} + s_{15}^2 s_{12} 
   + s_{12} s_{14} s_{35}
   \notag\\
   &+\frac{1}{2}s_{12}s_{14}s_{45} -s_{12}s_{15}s_{34}
   -\frac{1}{2} s_{12} s_{25} s_{45} -s_{13} s_{14} s_{35}
   -\frac{1}{3} s_{13} s_{15} s_{35} +2 s_{13} s_{24} s_{25}
   \notag\\
   &-s_{14} s_{23} s_{24}-s_{14} s_{23} s_{25}-s_{15} s_{23} s_{24}+s_{15} s_{24} s_{25}
   + \text{cycl} \Bigr)\,.\label{eq:A66bcj}
\end{align} 

\subsubsection*{Anomaly}
We have so far completely ignored the amplitudes descended from the anomalous sector. And for a good reason -- it is impossible to fulfill any of the above relations (either KK or BCJ) if the chiral anomaly is involved. This is even worse -- there is not even a trivial solution as the WZW Lagrangian~(\ref{eq:L5}) has no free parameter. Adding an extra constant in front of the WZW vertex would result in demand that this constant and all constants of the anomalous sector must be zero. This includes both the single anomaly insertion and the anomaly in the even amplitudes (for example, the anomalous part of the 8pt amplitude~(\ref{eq:A86aa})).
\subsection{String theory constraints}\label{sec:strings}
%
We ended up with an explicit form of the 6-pt amplitude at the $O(p^6)$ order, which satisfies the Adler zero condition and the BCJ constraints. That these constraints also hold for higher orders was verified explicitly for the 8-pt amplitude, and its form can be found in Appendix~\ref{sec:ap8pt}. Interestingly, the amplitudes depend only on one parameter (not counting the scale $F$). This is partly because the $O(p^4)$ order vanishes. Nevertheless, it is tempting to ask if our ``ChPT+BCJ'' theory can be reduced even more. If up to $O(p^6)$ we have only one dimensionful parameter $F$ and eventually one dimensionless parameter $C^{(6)}_1$ it means to set the latter parameter to some value. It would leave only one parameter for the whole theory. A similar situation is in the string theory, where there is only one dimensional constant -- the inverse of the string tension $\alpha'$ which is proportional to $\text{\it mass}^2$ or in our language: $\alpha' \sim 1/F^2$. The string theory in the low energy limit also describes scattering amplitudes of various quantum field theories. Even though this was typically aimed to describe gravity, recently in~\cite{Carrasco:2016ldy} it was also applied on obtaining (though indirectly) NLSM. Their work focuses on the $Z$-function, which plays a crucial role in the double copy for the open-string amplitudes~\cite{Broedel:2013tta}. They established a relation between the low energy limit of the $Z$-function and the color-ordered NLSM. Including also the higher-order corrections, a relevant amplitude for the given order and multiplicity is simply the coefficient in the $\alpha'$ expansion. We can summarize it by the following closed-form:\footnote{Note the sign, which is there due to using the mostly minus sign convention, our definitions of $s_{ij}$, the form of amplitudes and the high-energy behavior of $Z_\times$ in the physical region.}
\begin{equation}\label{eq:Astring}
    A^{(r)\text{string}}_n = -\alpha'^{\frac{r+n-4}{2}} \frac{1}{(r/2+n-3)!} Z_\times^{(r/2+n-3)}(\alpha'=0)\,,
\end{equation}
where $Z_\times^{(n)}$ is the $n$-the derivative of the $Z$-function with respect to $\alpha'$ evaluated in 0. The Abelian disk integrals $Z_\times$ are connected with the $Z$-functions via:
\begin{equation}
    Z_\times (p_1,\ldots,p_n) = \sum_{\sigma\in S_n/Z_n} Z_{\sigma_1\ldots \sigma_n}(p_1,\ldots,p_n)\,.
\end{equation}
The $Z$-functions are iterated integrals over the boundary of a disk worldsheet. The open-superstring amplitudes can be related as a double-copy of color-stripped Yang-Mills amplitudes and these $Z$-functions.

For the 4pt it is possible to write down a closed formula for all orders in $\alpha'$ using a sum of the three beta functions. Factoring out the Veneziano amplitude it takes the form:
\begin{equation}\label{eq:Zx4}
    Z_\times (1,2,3,4) = \Bigl(1 + \frac{\Gamma(1 + \alpha' u) \Gamma(-\alpha' u)}{\Gamma(1 + \alpha' s) \Gamma(-\alpha' s)}  + \frac{\Gamma(1 + \alpha' u) \Gamma(-\alpha' u)}{\Gamma(1 + \alpha' t) \Gamma(-\alpha' t)}  \Bigr) \frac{2}{-s} \frac{\Gamma(1 - \alpha' s) \Gamma(1 - \alpha' t)}{\Gamma(1 + \alpha' u)}.
\end{equation}
After demanding that $A^{(r)\text{BCJ}}_4$ in~(\ref{A4rbcj}) is equal to $A^{(r)\text{string}}_4$ obtained by plugging~(\ref{eq:Zx4}) into~(\ref{eq:Astring}), we have to fix the scales as:
\begin{equation}\label{eq:alphaF}
    \alpha' = \frac{1}{2\pi^2F^2}\,.
\end{equation}
The 4-pt amplitude up to all orders is then given by~(\ref{eq:Zx4}) (with $\alpha'$ replaced by $F^2$ using~(\ref{eq:alphaF})):
\begin{multline}
     A^{\text{string}}_4=\sum_{r=1}^\infty A^{(2r)\text{string}}_4 = -2 \pi^2 F^2 Z_\times\\ = -\frac{t}{2 F^2} - \frac{t(s^2+t^2+u^2)}{192\pi^2F^6} - \frac{\zeta(3)t(s^3+t^3+u^3)}{48\pi^6 F^8} +  \ldots\,,
\end{multline}
where we have also explicitly showed the terms up to $O(p^8)$. 
It is trivial to read off the 4pt constants. In our notation, as we left it in~(\ref{eq:kbcj}), starting with the known $k_1^{(2)} = -1/6$ and irrelevant $k_1^{(4)}$ (due to vanishing NLO), they are completely fixed by the $Z$-theory as:
\begin{align}
    \text{string:}\quad\, &k_1^{(6)} =-\frac{1}{2}\frac{\zeta(2)}{(2\pi)^4}= \frac{-1}{192\pi^2},\quad
    k_1^{(8)} = -\frac{4}{3}\frac{\zeta(3)}{(2 \pi)^6},\quad k_1^{(10)}= -3 \frac{\zeta(4)}{(2\pi)^8}= \frac{-1}{\num{7680} \pi^4}\notag\\
    &
    k_1^{(12)} =-\frac{8}{5}\frac{\zeta(2)\zeta(3)+2\zeta(5)}{(2\pi)^{10}} = -\frac{\pi^2 \zeta(3)+12 \zeta(5)}{3840 \pi^{10}},\notag\\
    &k_1^{(14)} = -\frac{51}{4}\frac{\zeta(6)}{(2\pi)^{12}}= -\frac{17}{\num{5160960}\,\pi^6},\notag\\
    &
    k_2^{(14)} =\frac{13\zeta(2)\zeta(4)- 16 \zeta(3)^2}{(2\pi)^{12}} =\frac{13 \pi^6-8640 \zeta(3)^2}{\num{2211840}\,\pi^{12}}\,.
\end{align}
Of course, if needed any higher order can be readily obtained. We can also easily re-express them in different parametrizations. For example the constants of Lagrangian~(\ref{eq:L6}) are
\begin{equation}
   \text{string:}\quad\;  C^{(6)}_1 = \frac{1}{768 \pi^2}\,,\quad
    C^{(6)}_2 = \frac{1}{1536 \pi^2}\,.
\end{equation}
This parametrization is important because we have expressed higher point amplitudes using the $C^{(6)}_1$. We have explicitly verified that our NNLO amplitude~(\ref{eq:A66bcj}) corresponds to the $Z$-theory result (cf.~(4.7) in \cite{Carrasco:2016ldy}) with the above mentioned $C^{(6)}_1$ value.

In the previous section, we have stopped at the $O(p^6)$ order for the form of the BCJ amplitudes with higher multiplicities (for $n>4$). We have, however, verified that the known form of the $O(p^8)$ six-point amplitude obtained from the corresponding subleading $\alpha'^7$ correction of $Z_\times$ (cf. Appendix~B in~\cite{Carrasco:2016ldy}) has the Adler zero and must be thus expressible by the ``ChPT+BCJ'' Lagrangian.  

\section{Summary and discussion}
In the first part of this work, we have summarized the canonical forms of the ChPT Lagrangian as constructed using the symmetric breaking pattern $H\times H\to H$ with unitary group $H$ of degree $N_f$. We have focused only on single-trace operators without external sources and chiral corrections, i.e., we work in the strict massless limit, in both even and odd intrinsic-parity sectors.
We have then studied the bottom-up construction of amplitudes in the respective sectors using the modern on-shell methods. We pushed the results to the following limits: for the 4pt scattering amplitudes to all orders, for multiplicity 6, 8, and 10 up to order $O(p^{10})$. In the odd sector, for the lowest multiplicity (5pt scatterings), it was given schematically again to all orders. We have also presented Feynman rules for the vertices of the Wess-Zumino-Witten Lagrangian (which is of the lowest $O(p^4)$ order) for all multiplicities in the so-called Cayley parametrization. Concerning the amplitudes in the odd sector, we got up to NNLO ($O(p^8)$) and calculated the seven-point scattering. For the first time, we presented and discussed the 8pt scattering with both -- even sector contributions and the anomalous one (i.e., with two insertions of the WZW vertices). In the so-called symmetric-center point, we have discussed the relevance of individual terms and concluded that the anomalous contribution is definitely non-negligible, at least without a deeper knowledge of other low-energy constants. As a by-product of the amplitude calculations, we have counted the number of independent monomials of relevant basis at given orders. We got an agreement with existing literature \cite{Bijnens:2018lez} and \cite{Dai:2020cpk} up to $O(p^8)$ order in number of terms for multiplicity $n=6$ and $n=8$ even-though in their respective works they considered a general dimension. We have used the bonus relations as a verification of our results. We have then focused on the $O(p^{10})$ order and discussed the difference from results obtained in~\cite{Dai:2020cpk}. We have determined and verified that the difference is due to the Gram determinant relations and the extra constraints among the monomials are thus of the Galileon-like form.

The second part involved a completely different point of view. We put further constraints on the amplitudes obtained in the first part, namely the Kleiss-Kuijf and BCJ relations. We found out it is possible to comply with them but only in the even sector. We have also verified that the KK relations are a subset of BCJ (i.e., whatever satisfies BCJ would also agree with KK). It is worth pointing out that the odd sector does not fit into this picture even if we add an arbitrary constant in front of the WZW vertex and study the behavior of two insertions of such a vertex (i.e., the even amplitude). The only immediate solution is to put such a constant equal to zero. It suggests that ChPT does not belong to the class of theories satisfying KK or/and BCJ. Another possibility is that we are missing some corrections to relations or some important contributions. For example, these contributions can come from the loop expansion (suggested by the $1/\pi^2$ in the WZW term). It would effectively mean that the anomaly is subleading and should be disregarded at the first order. If this is taken seriously, our obtained simplified effective theories, ``ChPT+KK'' or even more restrictive ``ChPT+BCJ'', might still represent QCD in some limit. 

We can continue even further and try to interconnect the string theory, or more precisely the so-called $Z$-theory, with our ``ChPT+BCJ'' amplitudes. Such cultivated theory is given only by one scale $\alpha'$ or $1/F^2$, and especially the 4pt scatterings are then known up to all orders in a closed form. We have explicitly verified that the formula is in agreement with existing calculations in the literature. We have also provided the expression for the $n=8$ amplitude within this theory, and it would be interesting to verify it from the direct $Z$-theory calculation using the eight-point disk integrals. Another possible direction is studying the status of our special amplitudes beyond the tree level; for example, comparing them with recent studies of one-loop six-point calculations \cite{Bijnens:2021hpq} or leading-logs evaluations \cite{Bijnens:2012hf,Bijnens:2013yca}. 

In conclusion, let us stress that the studied simplifications in the parametric space, and with these connected reduced theories, i.e. ``ChPT+KK'', ``ChPT+BCJ'' or finally ``ChPT+string'' should be taken with a grain of salt, especially if applied in the meson phenomenology. On the other hand, it would be interesting to see how they work in some suitable applications. 

\acknowledgments
I am grateful to Johan Bijnens, Tomas Husek, Jiri Novotny and Jaroslav Trnka for stimulating discussions and comments.
This work was supported by the Czech Government project GA \v{C}R 21-26574S. 

\appendix
\section{Complete set of single trace monomials at $O(p^8)$}\label{sec:apA}
We merely copied the relevant monomials from \cite{Bijnens:2018lez}, i.e. those with a single trace, without the mass or external-field insertion. As explained in the main text, now effectively $h_{\mu\nu} \circeq 2 \nabla_\mu u_\nu$, we prefer the symmetric $h_{\mu\nu}$ instead of $\nabla_\mu u_\nu$, which is dominantly used in \cite{Bijnens:2018lez}. On top of that a trivial manipulations with symmetric $h_{\mu\nu}$ were performed. Namely, the monomial
\begin{equation}
    O_4^{(8)} = \langle \nabla^\mu u^\nu \nabla_\nu u^\rho \nabla_\rho u^\sigma \nabla_\sigma u_\mu \rangle +
                \langle \nabla^\mu u^\nu \nabla^\rho u_\mu \nabla^\sigma u_\rho \nabla_\nu u_\sigma \rangle\,,
\end{equation}
defined in \cite{Bijnens:2018lez} can be easily rewritten as
\begin{equation}
    O_4^{(8)} \circeq \frac{1}{16} \langle h_{\mu\nu}h^{\nu\rho} h_{\rho\sigma} h^{\sigma\mu}\rangle + \frac{1}{16} \langle h_{\mu\nu}h^{\rho\mu} h_{\sigma\rho} h^{\nu\sigma}\rangle = \frac{1}{8} \langle h_{\mu\nu}h^{\nu\rho} h_{\rho\sigma} h^{\sigma\mu}\rangle \,,
\end{equation}
which is the first term in Tab.\ref{tab:op81}. Apart from the trivial manipulation and with this connected $1/2^n$ factors, the terms in Tabs.~\ref{tab:op81}, \ref{tab:op82}, \ref{tab:op83} correspond exactly to those in \cite{Bijnens:2018lez}.
\begin{table}[t]
\begin{center}  
\begin{tabular}{|c|c|}
\hline
 $i$ & monomials $\langle h^4 \rangle$ \\ \hline 
1 & $\langle h_{\mu\nu}h^{\nu\rho}h_{\rho\sigma}h^{\mu\sigma}  \rangle$  \\
\hline
2 & $\langle h_{\mu\nu}h^{\nu\rho}h^{\mu\sigma}h_{\rho\sigma}\rangle$  \\
\hline
3 & $\langle h_{\mu\nu}h_{\rho\sigma}h^{\mu\nu}h^{\rho\sigma}\rangle$  \\
\hline
    \end{tabular}
   \renewcommand{\arraystretch}{1}
     
    \caption{Independent single trace monomials at $O(p^8)$ starting with four pions  (corresponding to ${\cal O}^{(8)}_{4},\,{\cal O}^{(8)}_{5}$ and ${\cal O}^{(8)}_{6}$ of \cite{Bijnens:2018lez}).}\label{tab:op81}
  \end{center}
\end{table}

\begin{table}[t]
\begin{center}  
\begin{tabular}{|c|c|}
\hline
 $i$ & monomials $\langle h^2 u^4 \rangle$ \\ \hline 
4 & $\langle \{u^\mu u_\mu,\, u^\nu\} h^{\rho\sigma}u_\nu h_{\rho\sigma}\rangle$  \\
\hline
5 & $\langle u^\mu u^\nu u_\nu u_\mu h^{\rho\sigma}h_{\rho\sigma}\rangle$  \\
\hline
6 & $\langle u^\mu u^\nu u_\nu u_\rho h_{\mu\sigma}h^{\rho\sigma}\rangle$  \\
\hline
7 & $\langle u^\mu u^\nu u_\nu u_\rho h^{\rho\sigma}h_{\mu\sigma}\rangle$  \\
\hline
8 & $\langle u^\mu u^\nu h_{\mu\nu} u^\rho u^\sigma h_{\rho\sigma} + u^\mu u^\nu h_{\rho\sigma} u^\rho u^\sigma h_{\mu\nu}\rangle$  \\
\hline
9 & $\langle u^\mu u^\nu h_{\nu\rho} u^\rho u^\sigma h_{\mu\sigma}\rangle$  \\
\hline
10 & $\langle u^\mu u^\nu h_{\nu\rho} u^\sigma u^\rho h_{\mu\sigma}\rangle$  \\
\hline
11 & $\langle u^\mu u^\nu u^\rho h_{\mu\nu} u^\sigma h_{\sigma\rho} + u^\mu u^\nu u^\rho h_{\mu\sigma} u^\sigma h_{\nu\rho}\rangle$  \\
\hline
12 & $\langle u^\mu u^\nu u^\rho h_{\mu\rho} u^\sigma h_{\nu\sigma} + u^\mu u^\nu u^\rho h_{\nu\sigma} u^\sigma h_{\mu\rho} \rangle$  \\
\hline
13 & $\langle u^\mu u^\nu u^\rho h_{\nu\rho} u^\sigma h_{\mu\sigma} + u^\mu u^\nu u^\rho h_{\rho\sigma} u^\sigma h_{\mu\nu} \rangle$  \\
\hline
14 & $\langle u^\mu u^\nu u^\rho u^\sigma \{ h_{\mu \sigma},\, h_{\nu \rho}\}\rangle$  \\
\hline
15 & $\langle u^\mu u^\nu u^\rho u^\sigma h_{\nu \sigma} h_{\mu \rho} \rangle$  \\
\hline
16 & $\langle u^\mu u^\nu u^\rho u^\sigma h_{\rho\sigma} h_{\mu \nu} \rangle$  \\
\hline
17 & $\langle u^\mu u^\nu u_\rho h_{\mu\sigma}u_\nu h^{\rho\sigma}\rangle$  \\
\hline
18 & $\langle u^\mu u_\nu u_\rho h_{\mu\sigma}u^\rho h^{\nu\sigma} + u^\mu u^\nu u_\rho h_{\nu\sigma}u_\mu h^{\rho\sigma}\rangle$  \\
\hline
19 & $\langle u_\mu u^\nu u^\rho h_{\nu\sigma}u_\rho h^{\mu\sigma} + u^\mu u_\nu u^\rho h_{\rho\sigma}u_\mu h^{\nu\sigma}\rangle$  \\
\hline
20 & $\langle u^\mu u^\nu u_\rho h^{\rho\sigma}u_\nu h_{\mu\sigma}\rangle$  \\
\hline
21 & $\langle u^\mu u^\nu h_{\mu\rho}u_\nu u_\sigma h^{\rho\sigma}+ u^\mu u^\nu h^{\rho\sigma}u_\nu u_\sigma h_{\mu\rho}\rangle$  \\
\hline
22 & $\langle u^\mu u^\nu h_{\mu\rho} u_\sigma u_\nu h^{\rho\sigma} + u^\mu u^\nu h^{\rho\sigma} u_\mu u_\sigma h_{\nu\rho}\rangle$  \\
\hline
23 & $\langle u^\mu u^\nu h_{\nu\rho} u_\mu u_\sigma h^{\rho\sigma} + u^\mu u^\nu h^{\rho\sigma}u_\sigma u_\nu h_{\mu\rho}\rangle$  \\
\hline
24 & $\langle u^\mu u^\nu h_{\rho\sigma} u_\mu u_\nu h^{\rho\sigma} \rangle$  \\
\hline
25 & $\langle u^\mu u^\nu h_{\rho\sigma} u_\nu u_\mu h^{\rho\sigma} \rangle$  \\
\hline
    \end{tabular}
   \renewcommand{\arraystretch}{1}
     
    \caption{22 independent single trace monomials at $O(p^8)$ starting with six pions  (corresponding to ${\cal O}^{(8)}_{45}-{\cal O}^{(8)}_{66}$ of \cite{Bijnens:2018lez}).}\label{tab:op82}
  \end{center}
\end{table}

\begin{table}[t]
\begin{center}  
\begin{tabular}{|c|c|}
\hline
 $i$ & monomials $\langle u^6 \rangle$ \\ \hline 
26 & $\langle u^{\mu} u_{\mu} u^{\nu} u_{\nu} u^{\rho} u_{\rho} u^{\sigma} u_{\sigma} \rangle$  \\
\hline
27 & $\langle u^{\mu} u_{\mu} u^{\nu} u_{\nu} u^{\rho} u^{\sigma} u_{\rho} u_{\sigma}\rangle$  \\
\hline
28 & $\langle u^{\mu} u_{\mu} u^{\nu} u_{\nu} u^{\rho} u^{\sigma} u_{\sigma} u_{\rho}\rangle$  \\
\hline
29 & $\langle u^{\mu} u_{\mu} u^{\nu} u^{\rho} u_{\nu} u^{\sigma} u_{\rho} u_{\sigma} \rangle$  \\
\hline
30 & $\langle u^{\mu} u_{\mu} u^{\nu} u^{\rho} u_{\nu} u^{\sigma} u_{\sigma} u_{\rho}\rangle$  \\
\hline
31 & $\langle  u^{\mu} u_{\mu} u^{\nu} u^{\rho} u^{\sigma} u_{\nu} u_{\rho} u_{\sigma}  \rangle$  \\
\hline
32 & $\langle u^{\mu} u_{\mu} u^{\nu} u^{\rho} u^{\sigma} u_{\nu} u_{\sigma} u_{\rho} +  u^{\mu} u_{\mu} u^{\nu
} u^{\rho} u^{\sigma} u_{\rho} u_{\nu} u_{\sigma}\rangle$  \\
\hline
33 & $\langle u^{\mu} u_{\mu} u^{\nu} u^{\rho} u^{\sigma} u_{\rho} u_{\sigma} u_{\nu}\rangle$  \\
\hline
34 & $\langle u^{\mu} u_{\mu} u^{\nu} u^{\rho} u^{\sigma} u_{\sigma} u_{\nu} u_{\rho}\rangle$  \\
\hline
35 & $\langle u^{\mu} u_{\mu} u^{\nu} u^{\rho} u^{\sigma} u_{\sigma} u_{\rho} u_{\nu}\rangle$  \\
\hline
36 & $\langle u^{\mu} u^{\nu} u_{\mu} u_{\nu} u^{\rho} u^{\sigma} u_{\rho} u_{\sigma}\rangle$  \\
\hline
37 & $\langle u^{\mu} u^{\nu} u_{\mu} u^{\rho} u_{\nu} u^{\sigma} u_{\rho} u_{\sigma}\rangle$  \\
\hline
38 & $\langle  u^{\mu} u^{\nu} u_{\mu} u^{\rho} u^{\sigma} u_{\nu} u_{\rho} u_{\sigma} \rangle$  \\
\hline
39 & $\langle u^{\mu} u^{\nu} u_{\mu} u^{\rho} u^{\sigma} u_{\nu} u_{\sigma} u_{\rho}\rangle$  \\
\hline
40 & $\langle u^{\mu} u^{\nu} u^{\rho} u_{\mu} u^{\sigma} u_{\nu} u_{\rho} u_{\sigma}\rangle$  \\
\hline
41 & $\langle u^{\mu} u^{\nu} u^{\rho} u_{\mu} u^{\sigma} u_{\rho} u_{\nu} u_{\sigma} \rangle$  \\
\hline
42 & $\langle u^{\mu} u^{\nu} u^{\rho} u^{\sigma} u_{\mu} u_{\nu} u_{\rho} u_{\sigma} \rangle$  \\
\hline
    \end{tabular}
   \renewcommand{\arraystretch}{1}
     
    \caption{17 independent single trace monomials at $O(p^8)$ starting with eight pions (equal to ${\cal O}^{(8)}_{119}-{\cal O}^{(8)}_{135}$ of \cite{Bijnens:2018lez}).}\label{tab:op83}
  \end{center}
\end{table}

\section{Cayley parametrization}\label{sec:cayley}
It is known that the $U(1)$ piece of $u(\phi)$ in LO (\ref{eq:L2}) decouples \cite{Kampf:2013vha}. It means that including also a normalized identity matrix for $a=0$ in (\ref{eq:udef}) with $t^0 = I_N/\sqrt{N}$ any amplitude with at least one $\phi^0$ is zero. It is true only for the leading order Lagrangian (\ref{eq:L2}) and not e.g. for ${\cal L}_4$ or higher orders. However, it is interesting to notice that it is again true for the leading odd-intrinsic ${\cal L}_5$ given in (\ref{eq:L5}). We can thus use the advantages of the bigger symmetry group $U(N)$, and some results could then be particularly simpler.

We will define the Cayley parametrization of the $U(N)$ non-linear sigma model as
\begin{equation}
    U = \frac{1 + \frac{i}{\sqrt2 F} \Phi}{1 - \frac{i}{\sqrt2 F} \Phi}\,.
\end{equation}
Note that our definition of $\Phi = t^a \phi^a$ (cf.~(\ref{eq:udef})). Within this parametrization the stripped Feynman rule for the interaction vertices is remarkably simple \cite{Kampf:2013vha}
\begin{equation}
    V_{2n}  = \Bigl(-\frac{1}{2 F^2}\Bigr)^{n-1} \left(\sum_{i=0}^{n-1} p_{2i+1}\right)^2\,,
\end{equation}
so for example the 4pt LO vertex is given by
\begin{equation}\label{eq:V4}
    V_4 = -\frac{1}{2 F^2} (p_1 + p_3)^2 = -\frac{1}{2 F^2} (p_2 + p_4)^2\,.
\end{equation}
Now we will derive a similar Feynman rule for the WZW Lagrangian~(\ref{eq:L5}). The coset corresponding to the exponential parametrization~(\ref{eq:Uz}) is simply
\begin{equation}
    {\cal U} = \frac{1 + \frac{i}{\sqrt2 F} z \Phi}{1 - \frac{i}{\sqrt2 F}z \Phi}\,.
\end{equation}
The five-dimensional quantity $\Sigma_i$ is then given by
\begin{align}
    &\Sigma_z = {\cal U}^\dagger\partial_z {\cal U} = \frac{i\sqrt{2}}{F} \Phi \frac{1}{1+z^2\frac{\Phi^2}{2F^2}}\notag\\
    &\Sigma_\mu = {\cal U}^\dagger\partial_\mu {\cal U} = \frac{i\sqrt{2}}{F} z \partial_\mu\Phi \frac{1}{1+z^2\frac{\Phi^2}{2F^2}}\,.
\end{align}
Plugging back into ${\cal L}_5$ given by~(\ref{eq:L5}) we can easily calculate a vertex for any given (odd) number of pions by expanding several of $1/(1+z^2\Phi^2/2F^2)$ to the appropriate order. We will now show how to obtain the closed-form for a given number of fields. First, we take advantage of the integral in (\ref{eq:L5}) and via a change of variable
\begin{equation}
    z = i \tilde z \sqrt{2}F\,,
\end{equation}
we get a correct sign for the geometrical series of the denominator expansion. The Lagrangian becomes
\begin{equation}\label{eq:L5expanded}
    {\cal L}_5 = i\frac{2 N_C}{3\pi^2} \int_0^{1/i\sqrt2F}dz z^4 \epsilon^{\mu\nu\alpha\beta} \Bigl\langle \partial_\mu \Phi \frac{1}{1-z^2 \Phi^2} \partial_\nu \Phi \frac{1}{1-z^2 \Phi^2} \partial_\alpha \Phi \frac{1}{1-z^2 \Phi^2} \partial_\beta \Phi \frac{\Phi}{(1-z^2 \Phi^2)^2}\Bigr\rangle
\end{equation}
The integration of $z$ is trivial -- it effectively brings $z^n/n$ for the $n$-pion interaction. The geometrical expansion of $1/(1-z^2\Phi^2)$ is as well trivial and the only non-trivial (but still very simple) factor can come from the last term in the trace. The reduced Feynman rule can be thus directly read off from~(\ref{eq:L5expanded}) to take the following form
\begin{equation}
    V_n^{WZW} = \frac{i^{n-1}}{n}\frac{2 N_C}{3\pi^2(\sqrt{2}F)^n}
    \sum_{i=0}^{\frac{n-5}{2}}
    \sum_{j=i}^{\frac{n-5}{2}}
    \sum_{k=j}^{\frac{n-5}{2}}
    \epsilon^{p_1\,p_{2+2i}\,p_{3+2j}\,p_{4+2k}}\Bigl(\frac{n-3}{2} -k \Bigr) + \text{cycl}\,,
\end{equation}
for all odd $n\geq5$, where we sum over all cyclic permutations in momenta. We have verified this expression up to $n=9$ using the direct calculation. In the main text we will need the explicit form of the following 5- and 7-point vertices:
\begin{align}\label{eq:V5}
    &V_5 = \frac{N_C}{6\sqrt2\pi^2 F^5} \epsilon^{p_1\,p_2\,p_3\,p_4}\\\label{eq:V7}
    &V_7 = \frac{-N_C}{12\sqrt2\pi^2 F^7} (\epsilon^{p_1\,p_2\,p_3\,p_4} +\epsilon^{p_1\,p_2\,p_3\,p_6} +\epsilon^{p_1\,p_2\,p_5\,p_6} +\epsilon^{p_1\,p_4\,p_5\,p_6} +\epsilon^{p_3\,p_4\,p_5\,p_6} )\,.
\end{align}

\section{8-pt BCJ amplitudes}\label{sec:ap8pt}
The tree-level 8pt amplitudes up to $O(p^6)$, which satisfy both the Adler zero and BCJ conditions, will be summarized here. For the following results, we have used the so-called minimal parametrization discussed in~\cite{Kampf:2013vha} and variables $X_{ij}$ defined in~(\ref{eq:Xij}). 

We start with the LO. As we know, any $O(p^2)$ amplitude in NLSM automatically satisfies also BCJ. The minimal parametrization seems to lead to the most economical form at $O(p^2)$, and our amplitude is indeed relatively short:
\begin{align}
    F^{6} &A^{(2)}_8 =  \frac{(X_{13}+X_{24}) (X_{15}+X_{48}) (X_{57}+X_{68})}{16 X_{14}X_{58}} + \frac{(X_{13}+X_{24}) (X_{15}+X_{46}) (X_{17}+X_{68})}{8X_{14}X_{16}}\notag\\
    &+\frac{(X_{13}+X_{24}) (X_{15}+X_{17}+X_{46}+X_{48}+X_{57}+X_{68})}{8X_{14}}
    + \frac14 X_{13} + \frac{1}{16}X_{15} + \text{cycl}\,.
\end{align}
The NLO amplitude is zero, so we continue with the $O(p^6)$ order.
It is a quite lengthy expression but similar in structure to the $O(p^2)$: with parts with double propagators, single propagators and contact terms. It would be interesting to try to find the shortest possible form, similar to the 6pt amplitude given in~(\ref{eq:A66bcj}).  The double and single factorization terms are explicitly given by:
\allowdisplaybreaks[2]
\begin{align}
   F^{10} &A^{(6)\text{BCJ}}_8 = 
   \frac{2 C^{(6)}_1}{X_{14} X_{16}}(X_{13}+X_{24}) (X_{15}+X_{46}) (X_{17}+X_{68}) (X_{13}^2+X_{15}^2+X_{17}^2\notag\\
   &\hspace{70pt}+X_{24}^2 +X_{46}^2+X_{68}^2 + X_{13} X_{24} +X_{15} X_{46}+X_{17}
   X_{68})
   \notag\\
   &+\frac{C^{(6)}_1}{X_{14} X_{58}}(X_{13}+X_{24})(X_{15}+X_{48})(X_{57}+X_{68})
   (X_{13}^2+X_{15}^2+X_{24}^2\notag\\
   &\hspace{20pt}+X_{48}^2+X_{57}^2+X_{68}^2 +X_{13} X_{24}+X_{15} X_{48}+X_{57}
   X_{68})
   \notag\\
   &-\frac{2 C^{(6)}_1(X_{13}+X_{24})}{X_{14}} \bigl(
   X_{15}^3+X_{17}^3+X_{46}^3+X_{48}^3+X_{57}^3+X_{68}^3
   \notag\\&\hspace{20pt}
   +X_{13}^2 X_{15}+X_{13}^2 X_{17}+X_{13}^2 X_{46}+X_{13}^2 X_{48}+X_{13}^2 X_{57}+X_{13}^2 X_{68}+X_{15} X_{24}^2
   \notag\\&\hspace{20pt}
   -2 X_{15} X_{47}^2+2 X_{15} X_{68}^2+2 X_{15}^2 X_{47}+2 X_{15}^2 X_{68}+2 X_{16} X_{47}^2+2 X_{16} X_{48}^2+2 X_{16} X_{57}^2
   \notag\\&\hspace{20pt}
   +2 X_{16} X_{58}^2+2 X_{16}^2 X_{47}-2 X_{16}^2 X_{48}-2 X_{16}^2 X_{57}+2 X_{16}^2 X_{58}+X_{17} X_{24}^2+2 X_{17} X_{46}^2
   \notag\\&\hspace{20pt}
   -2 X_{17} X_{58}^2+2 X_{17}^2 X_{46}+2 X_{17}^2 X_{58}+X_{24}^2 X_{46}+X_{24}^2 X_{48}+X_{24}^2 X_{57}+X_{24}^2 X_{68}
   \notag\\&\hspace{20pt}
   -2 X_{46} X_{58}^2+2 X_{46}^2 X_{58}+2 X_{47} X_{58}^2+2 X_{47} X_{68}^2+2 X_{47}^2 X_{58}-2 X_{47}^2 X_{68}+2 X_{48} X_{57}^2
   \notag\\&\hspace{20pt}
   +2 X_{48}^2 X_{57}
   + X_{15} X_{16} X_{17}+X_{13} X_{24} X_{17}+2 X_{15} X_{46} X_{17}-X_{16} X_{46} X_{17}
   \notag\\&\hspace{20pt}
   +2 X_{15} X_{47} X_{17}-2 X_{16} X_{47} X_{17}-X_{46} X_{47} X_{17}-4 X_{15} X_{48} X_{17}+4 X_{16} X_{48} X_{17}
   \notag\\&\hspace{20pt}
   +2 X_{46} X_{48} X_{17}+X_{15} X_{57} X_{17}+2 X_{16} X_{57} X_{17}+2 X_{46} X_{57} X_{17}+X_{47} X_{57} X_{17}
   \notag\\&\hspace{20pt}
   +2 X_{48} X_{57} X_{17}+2 X_{15} X_{58} X_{17}-4 X_{16} X_{58} X_{17}-4 X_{47} X_{58} X_{17}+4 X_{48} X_{58} X_{17}
   \notag\\&\hspace{20pt}
   +2 X_{57} X_{58} X_{17}+2 X_{15} X_{68} X_{17}+2 X_{46} X_{68} X_{17}+4 X_{47} X_{68} X_{17}-4 X_{48} X_{68} X_{17}
   \notag\\&\hspace{20pt}
   -4 X_{57} X_{68} X_{17}+4 X_{58} X_{68} X_{17}+X_{13} X_{15} X_{24}+X_{13} X_{24} X_{46}-4 X_{15} X_{16} X_{47}
   \notag\\&\hspace{20pt}
   +4 X_{15} X_{46} X_{47}-2 X_{16} X_{46} X_{47}+4 X_{15} X_{16} X_{48}+X_{13} X_{24} X_{48}-4 X_{15} X_{46} X_{48}
   \notag\\&\hspace{20pt}
   +2 X_{16} X_{46} X_{48}+4 X_{15} X_{47} X_{48}-4 X_{16} X_{47} X_{48}+X_{46} X_{47} X_{48}+2 X_{15} X_{16} X_{57}
   \notag\\&\hspace{20pt}
   +X_{13} X_{24} X_{57}-4 X_{15} X_{46} X_{57}+4 X_{16} X_{46} X_{57}+2 X_{15} X_{47} X_{57}-4 X_{16} X_{47} X_{57}
   \notag\\&\hspace{20pt}
   +2 X_{15} X_{48} X_{57}+2 X_{46} X_{48} X_{57}-X_{47} X_{48} X_{57}-2 X_{15} X_{16} X_{58}+4 X_{15} X_{46} X_{58}
   \notag\\&\hspace{20pt}
   -4 X_{16} X_{46} X_{58}-4 X_{15} X_{47} X_{58}+8 X_{16} X_{47} X_{58}-4 X_{46} X_{47} X_{58}-4 X_{16} X_{48} X_{58}
   \notag\\&\hspace{20pt}
   +2 X_{46} X_{48} X_{58}-2 X_{47} X_{48} X_{58}+X_{15} X_{57} X_{58}-4 X_{16} X_{57} X_{58}+4 X_{46} X_{57} X_{58}
   \notag\\&\hspace{20pt}
   -2 X_{47} X_{57} X_{58}-X_{48} X_{57} X_{58}-X_{15} X_{16} X_{68}+X_{13} X_{24} X_{68}+2 X_{15} X_{46} X_{68}
   \notag\\&\hspace{20pt}
   +X_{16} X_{46} X_{68}-4 X_{16} X_{47} X_{68}+2 X_{46} X_{47} X_{68}+2 X_{15} X_{48} X_{68}+2 X_{16} X_{48} X_{68}
   \notag\\&\hspace{20pt}
   +X_{46} X_{48} X_{68}+2 X_{47} X_{48} X_{68}+2 X_{15} X_{57} X_{68}+4 X_{16} X_{57} X_{68}-4 X_{46} X_{57} X_{68}
   \notag\\&\hspace{20pt}
   +4 X_{47} X_{57} X_{68}+2 X_{48} X_{57} X_{68}-X_{15} X_{58} X_{68}-2 X_{16} X_{58} X_{68}+2 X_{46} X_{58} X_{68}
   \notag\\&\hspace{20pt}
   -4 X_{47} X_{58} X_{68}+X_{48} X_{58} X_{68}\bigr) + \text{cycl} + \text{cont.}\,.
\end{align}
The Adler zero condition can unambiguously set the contact terms as the 8pt amplitude at the $O(p^6)$ order is completely reconstructible using its factorizations. 

Both amplitudes depend only on $F$ and at $O(p^6)$ on one additional dimensionless constant. We can make them to correspond to the $Z$-theory (more precisely to the relevant term in the expansion in $\alpha'$) by setting $F^2= (2\pi^2\alpha')^{-1}$ and $C^{(6)}_1 = 1/768\pi^2$ (cf. Section~\ref{sec:strings}).

\bibliographystyle{myJHEP}
\bibliography{references}

\end{document}